\documentclass{article} 
\usepackage[left = 2.5cm, right = 2.5cm, bottom = 3cm, top = 2.5cm]{geometry}

\usepackage{amsmath,amssymb,amsthm,bm,bbm,dsfont,subcaption,float,tikz,array, booktabs, adjustbox}

\DeclareMathOperator{\expect}{E}
\DeclareMathOperator{\var}{var}

\DeclareMathOperator{\trace}{tr}

\DeclareMathOperator{\vect}{vec}
\DeclareMathOperator{\normal}{N}

\usepackage{natbib}
\usepackage{hyperref}
\usepackage{xr-hyper}
\usepackage{autonum} 
\usepackage{refcount, calc}

\usepackage{pdfpages}

\usepackage{xcolor}

\usepackage{enumitem}
\usepackage[framemethod=TikZ]{mdframed}

\newtheorem{theorem}{Theorem}[section]
\newtheorem*{theorem*}{Theorem}

\theoremstyle{definition}

\makeatletter
\newcommand{\vnorm}[1]{%
	\ensuremath{%
		\if@display
		\left\| #1 \right\|
		\else
		\| #1 \|
		\fi
	}%
}
\makeatother
\newcommand{\mnorm}[1]{{\left\vert\kern-0.25ex\left\vert\kern-0.25ex\left\vert #1 
		\right\vert\kern-0.25ex\right\vert\kern-0.25ex\right\vert}}
\newcommand{\mnorms}[1]{{\vert\kern-0.25ex\vert\kern-0.25ex\vert #1 
		\vert\kern-0.25ex\vert\kern-0.25ex\vert}}
\newcommand*{\bb}{\boldsymbol}

\def\bLambda{\bb{\Lambda}}
\def\blambda{\bb{\lambda}}
\def\bL{\bb{L}}
\def\bQ{\bb{Q}}

\def\bx{\bb{x}}
\def\bmu{\bb{\mu}}
\def\b0{\bb{0}}

\def\bA{\bb{A}}

\def\bC{\bb{C}}
\def\bS{\bb{S}}

\def\bSigma{\bb{\Sigma}}
\def\bTheta{\bb{\Theta}}

\def\btheta{\bb{\theta}}
\def\bz{\bb{z}}
\def\be{\bb{\epsilon}}
\def\bPsi{\bb{\Psi}} 
\def\bI{\bb{I}}
\def\bthat{\hat{\bb{\theta}}}
\def\bttilde{\tilde{\bb{\theta}}}
\def\btnod{\btheta_{0}}

\newcolumntype{C}{>{$}c<{$}}
\newcolumntype{R}{>{$}r<{$}} 

\usepackage{orcidlink}
\usepackage{authblk} % Package for formatting author and affiliation

\title{Maximum softly penalised likelihood in factor analysis}
\author[1]{Philipp Sterzinger\orcidlink{0000-0002-0001-2345} \thanks{Correspondence concerning this article should be addressed to Philipp Sterzinger, Department of Statistics, London School of Economics and Political Science. E-mail: \href{mailto:p.sterzinger@lse.ac.uk}{p.sterzinger@lse.ac.uk}}}
\author[2]{Ioannis Kosmidis\orcidlink{0000-0003-0002-3456}}
\author[1]{Irini Moustaki\orcidlink{0000-0001-8371-1251}}

\affil[1]{London School of Economics and Political Science, Department of Statistics}
\affil[2]{University of Warwick, Department of Statistics}

\begin{document}
\maketitle
	
\begin{abstract}
  Estimation in exploratory factor analysis often yields estimates on
  the boundary of the parameter space. Such occurrences, known as
  Heywood cases, are characterised by non-positive variance estimates
  and can cause issues in numerical optimisation procedures or
  convergence failures, which, in turn, can lead to misleading
  inferences, particularly regarding factor scores and model
  selection. We derive sufficient conditions on the model and a
  penalty to the log-likelihood function that i) guarantee the
  existence of maximum penalised likelihood estimates in the interior
  of the parameter space, and ii) ensure that the corresponding
  estimators possess the desirable asymptotic properties expected by
  the maximum likelihood estimator, namely consistency and asymptotic
  normality. Consistency and asymptotic normality are achieved when
  the penalisation is soft enough, in a way that adapts to the
  information accumulation about the model parameters. We formally
  show, for the first time, that the penalties of \citet{akaike:87}
  and \citet{hirose+etal:2011} to the log-likelihood of the normal
  linear factor model satisfy the conditions for existence, and,
  hence, deal with Heywood cases. Their vanilla versions, though, can
  result in questionable finite-sample properties in estimation,
  inference, and model selection. The maximum softly-penalised
  likelihood framework we introduce enables the careful scaling of
  those penalties to ensure that the resulting estimation and
  inference procedures inherit the ML estimator’s optimal properties. Through
  comprehensive simulation studies and the analysis of real data sets,
  we illustrate the desirable finite-sample properties of the maximum
  softly penalised likelihood estimators and associated procedures.
\end{abstract}

\section{Introduction} 

Exploratory factor analysis has been widely used in social sciences
and beyond to measure unobserved constructs of interest such as
ability, attitudes, and behaviours, and for dimensionality
reduction. It has been noted early on in the factor analysis literature,
particularly with the development of the more precise computational
frameworks for maximum likelihood (ML) estimation in
\citet{joreskog:67} and \citet{joreskog+lawley:68}, that the
estimation of factor analysis models often results in improper
solutions. Such improper solutions involve zero or negative estimates
for error variances, and often correlation estimates greater than
one in absolute value. Such occurrences are typically referred to as
Heywood cases \citep{heywood:31}.  \citet{martin.mcdonald:75}
distinguish between an exact Heywood case in which at least one of the
estimates of the error variances is zero but none are negative, and an
ultra-Heywood case, in which at least one estimate of the error
variances is negative. A zero error variance implies that there is no
measurement error, and the factors fully explain the observed
variable. That is rare in real applications but, at the same time,
does not pose as much concern as negative estimates for
error variances do. Causes of Heywood cases that have been reported in
the literature are model and data dependent and include outliers,
non-convergence of associated optimisation procedures,
under-identification, model misspecification, missing data, and
sampling fluctuations combined with a true value close to the boundary
for the parameter, small sample sizes, poorly defined factors, and
factor over-extraction (\citealp[see, for example,][and references
therein]{vandriel:78,dillon:87,kano:98,chen+etal:2001,cooperman.waller:22}).
\citet{cooperman.waller:22} provides an up-to-date review of the
causes, effects, and solutions to Heywood cases in confirmatory and
exploratory factor analysis.

The presence of Heywood cases in factor analysis has practical
implications. It can produce parameter estimates, standard errors,
factor scores, and goodness-of-fit test statistics that cannot be
trusted. \citet{cooperman.waller:22} found, through a simulation
study, that Heywood cases increase the standard errors of factor
loadings and bias the factor scores upwards. Eliminating items that
correspond to estimates which display a Heywood case often moves the
Heywood problem to one of the remaining items.

An approach to handle Heywood cases, especially when they are
suspected to be due to sampling fluctuations, is by restricting the
estimates of the error variances to $[0, \infty)$ either explicitly or
by setting negative estimates to zero; see
\citet{gerbing+anderson:1987} for a discussion. However, this violates
regularity conditions of maximum likelihood estimation, leading to
estimators and testing procedures with properties that are hard to
evaluate. Another common approach is to impose priors on the loadings,
error variances, or both to avoid improper solutions. Estimation,
then, proceeds either using a likelihood-based approach with the prior
information incorporated via a penalty term (\citealp[see, for
example,][]{martin.mcdonald:75,akaike:87,hirose+etal:2011,lee:81}) or
by posterior sampling through MCMC (\citealp[see, for
example,][]{lee.song:02}). For example, \citet{martin.mcdonald:75}
proposed a Bayesian estimation framework in which they maximise not
the likelihood but the posterior density, using a prior distribution
for the error covariance matrix that assigns zero probability to
negative values. They assume a prior distribution that is almost
uniform, except that it decreases to zero at the point where an error
variance is equal to zero. \citet{lee:81} also investigated the form
of the posterior density under different informative prior
distributions, some of which have been designed to deal with Heywood
cases. \citet{akaike:87}, in the process of developing a model
selection criterion for factor analysis, also encountered the problem
of improper solutions and proposed a standard spherical prior
distribution of factor loadings and a uniform distribution for the
error variances. \citet{hirose+etal:2011} build on Akaike's work by
imposing a prior distribution only on the error variances, where the
inverse of the diagonal elements of the error covariance matrix have
exponential distributions. To our knowledge, there has been no formal
proof that such penalties prevent Heywood cases. Furthermore, naive
penalisation can introduce considerable finite sample bias in the
estimation of the factor loadings and error variances, as illustrated
in the simulation studies of Section~\ref{sec:fasimuls}.

This paper introduces a maximum softly penalised likelihood (MSPL)
framework for factor models. Specifically, we derive sufficient
conditions on an arbitrary penalty to the log-likelihood function that
guarantee that maximum penalised likelihood (MPL) estimation never
results in the occurrence of Heywood cases. Furthermore, we show that
the penalties proposed in \citet{akaike:87} and
\citet{hirose+etal:2011} satisfy those conditions, while guaranteeing
key equivariance properties for factor analysis, namely equivariance
under arbitrary scaling of the data and under factor rotations. To our
knowledge, this is the first proof that those two penalties can
effectively deal with Heywood cases. We, then, present conditions,
under which MPL estimators have the desirable asymptotic properties
expected from the ML estimator, namely consistency and asymptotic
normality, under the assumption that the limit of the estimator of the
variance-covariance matrix decomposes in a exploratory factor analysis
fashion. In particular, we show that this is achieved by requiring
that the penalisation is soft enough, in a way that adapts to how
information about the model parameters accumulates. We also discuss
how the \citet{akaike:87} and \citet{hirose+etal:2011} penalties can
be adapted for soft penalisation.  Although our asymptotic results
assume correct specification, in which case Heywood cases are sampling
artefacts, our simulation studies about model selection performance in
Section~\ref{sec:fasimuls} suggest that under factor number
misspecification, MSPL estimation continues to exhibit the least-false
behaviour expected by the ML estimator \citep[see, for
example,][]{white:1982}.

The remainder of this paper is organised as
follows. Section~\ref{sec:fapl_model} briefly presents the factor
analysis model. The proposed MPL framework is introduced in
Section~\ref{sec:faplmspl}. Section~\ref{sec:fapl_existence} states
our existence result of MPL estimates, which rules out the occurrence
of Heywood cases, and Section~\ref{sec:fapl_asymp} provides the
asymptotic behaviour of MPL estimators. Section~\ref{sec:mspl}
discusses the scaling factors for our MSPL estimators, and
Section~\ref{sec:fasimuls} provides a series of simulation studies
that investigate the finite sample performance of MSPL-based
estimation and inference, and compares them to existing Bayesian
approaches. Section~\ref{sec:fapl_examples} gives real data examples
and final remarks are provided in Section~\ref{sec:fapl_sum}. Proofs
of all theoretical results and additional materials are provided in
the supplementary material document.

\section{Exploratory factor analysis}
\label{sec:fapl_model}

The factor analysis model for a random vector of observed variables $\bx$ and $q$ factors $(q < p)$ is
\begin{equation}
  \label{eq:fa_model}
  \bx = \bmu + \bLambda \bz +\be \,,
\end{equation}
where $\bmu = (\mu_1, \ldots, \mu_p)^\top \in \Re^p$, $\bLambda$ is a
$p \times q$ real matrix of factor loadings,
$\bz \sim \normal(\bb{0}_q, \bI_q)$,
$\be \sim \normal(\bb{0}_p, \bPsi)$, and $\bz$ is independent of
$\be$. In the latter expressions, $\bPsi$ is a $p \times p$ diagonal
matrix with $j$th diagonal element $\psi_{j} > 0$, and $\bb{0}_q$ is a
vector of $q$ zeros, and $\bI_q$ is the $q \times q$ identity
matrix. It follows that $\expect(\bx) = \bmu$ and
$\var (\bx) = \bSigma = \bLambda\bLambda^\top + \bPsi$. The
exploratory factor analysis model is identifiable only up to
orthogonal rotations of the factor loadings matrix $\bLambda$.
\citet[Chapter~3]{bartholomew.ea:11} discuss approaches that resolve
unidentifiability.

In the presence of realisations of $n$ independent random vectors
$\bx_1, \ldots, \bx_n$, the log-likelihood function about the
parameters $\bmu$ and $\bSigma$ of the exploratory factor analysis
model is
\begin{equation}
  \label{eq:loglikl}
    C - \frac{n}{2} \left[ \log \det \left( \bSigma \right) + \trace \left(\bSigma^{-1} \bS \right) +   \sum_{i=1}^{n} ( \bar{\bx} - \bmu)^\top \bSigma^{-1} (\bar{\bx} - \bmu)\right] \,,
\end{equation}
where $C = -np \log(2 \pi)/2$, $\bar{\bx} = \sum_{i=1}^n \bx_i /n$ and
$\bS = \sum_{i=1}^n (\bx_i - \bar{\bx})(\bx_i -\bar{\bx} )^\top / n$
is the sample covariance matrix, assumed to be full rank. Clearly, the
maximiser of~\eqref{eq:loglikl} with respect to $\bmu$ is $\bar{\bx}$,
and at that point the quadratic term in~(\ref{eq:loglikl}) involving
$\bar{\bx}$ and $\bmu$ vanishes. Then, the profile log-likelihood
about $\bPsi$ and $\bLambda$ is
\begin{equation}
  \label{eq:loglikl2}
  \ell(\btheta; \bS) = C - \frac{n}{2} \left[ \log \det \left( \bLambda \bLambda^\top + \bPsi \right) + \trace \left\{ \left(\bLambda\bLambda^\top + \bPsi\right)^{-1} \bS \right\} \right] \,,
\end{equation}
with
$\btheta = (\theta_1, \ldots, \theta_{p(q + 1)})^\top = (\lambda_{11},
\ldots, \lambda_{pq}, \psi_{1}, \ldots, \psi_{p})^\top$, where
$\lambda_{jk}$ and $\psi_{j}$ are the $(j, k)$th and $(j, j)$th
elements of $\bLambda$ and $\bPsi$, respectively
$(j = 1, \ldots, p; k = 1, \ldots, q)$. Heywood cases correspond to
directions $\{\btheta(t)\}_{t \in \mathbb{\Re}}$ such that the value
of $\ell(\btheta(t); \bS)$ increases but
$\lim\limits_{t \to \infty }\bPsi(\btheta(t))$ is no longer positive
definite. Ultra-Heywood cases, i.e.~maxima of \eqref{eq:loglikl2} with
some of the variances negative, can, of course, be prevented by
maximising the log-likelihood under the constraint that
$\psi_{j} > 0$. Nevertheless, this does not eliminate the possibility
of at least one of the maximum likelihood estimates of
$\psi_{11}, \ldots, \psi_{pp}$ being exactly zero.

\section{Maximum penalised likelihood for handling Heywood cases}
\label{sec:faplmspl}

A straightforward way to avoid Heywood cases is to employ an MPL
estimator
\begin{equation}
  \label{eq:fapl_MPL}
  \tilde{\btheta} \in \arg \underset{\btheta \in \bTheta}{\max} \, \ell^*(\btheta ; \bS) \,,
\end{equation}
where $\ell^*(\btheta; \bS) = \ell(\btheta; \bS) + P^*(\btheta; \bS)$ and
$\bTheta = \left\{\btheta \in \Re^{p (q + 1)}: \theta_m > 0, m > pq \right\}$, with a penalty function $P^*(\btheta)$ that discourages
estimates of $\psi_{ii}$ being zero, so that the set 
$$
\arg \underset{\btheta \in \bTheta}{\max} \, \ell^*(\btheta ; \bS) = \left\{\bar{\btheta}\in \bTheta: \ell^*(\bar \btheta; \bS) = \underset{\btheta \in \bTheta}{\sup} \, \ell^*(\btheta; \bS) \right\}
$$
is nonempty. Towards constructing a MPL estimator that always exists,
\citet{akaike:87} and \citet{hirose+etal:2011} proposed the penalties
\begin{equation}
  \label{eq:hirose}
  P^*(\btheta) = -\frac{\rho n}{2} \trace\left(\bPsi^{-1/2} \bLambda \bLambda^\top \bPsi^{-1/2}\right) \,,  \quad \text{and} \quad 
  P^*(\btheta) = -\frac{\rho n}{2} \trace\left(\bPsi^{-1/2} \bS \bPsi^{-1/2}\right) \,,
\end{equation}
respectively, for $\rho > 0$. The penalties in~(\ref{eq:hirose}) are
attractive because the MPL estimates preserve two particular
equivariance properties that the ML estimator has, namely equivariance
under rescaling of the response vectors and equivariance under
rotations of the factor loadings. The former is desirable to justify
the common practice in factor analysis of setting $\bS$
in~(\ref{eq:loglikl2}) to the sample correlation matrix, and the latter
is desirable because it ensures that any post-fit rotation of the
factors is still the ML estimate of the rotated factors.

To see the equivariance under rescaling of the response vectors, let
$\dot{\bx}_i = \bL \bx_i$, for a known, diagonal, invertible
$p \times p$ matrix $\bL$. Then,
$\dot{\bSigma} = \var(\dot{\bx}_i) = \dot{\bLambda}
\dot{\bLambda}^\top + \dot{\bPsi}$, with
$\dot{\bLambda} = \bL \bLambda$ and
$\dot{\bPsi} = \bL \bPsi \bL^\top$, and the sample variance-covariance
matrix based on $\dot{\bx}_1, \ldots, \dot{\bx}_n$ is
$\dot{\bS} = \bL \bS \bL^\top$. Denoting
$\dot{\btheta} = (\dot\lambda_{11}, \ldots, \dot\lambda_{pq},
\dot\psi_{1}, \ldots, \dot\psi_{p})^\top$, the cyclic property of
the trace operator and properties of the determinant for products of
invertible matrices can be used to show that
$\ell(\dot\btheta; \dot{\bS}) = \ell(\btheta; \bS) + \dot{c}$ where
$\dot{c}$ does not depend on $\dot\btheta$. Hence, if $\hat\bLambda$
and $\hat\bPsi$ are the maximisers of $\ell(\btheta; \bS)$, the
maximisers of $\ell(\dot\btheta; \dot{\bS})$ are $\bL \hat{\bLambda}$
and $\bL \hat{\bPsi} \bL^\top$, respectively. Similar calculations
show that, for both penalties in~(\ref{eq:hirose}),
$P^*(\dot\btheta) = P^*(\btheta) + \dot{d}$ for a known constant
$\dot{d}$ that does not depend on $\dot\btheta$. Hence, if
$\tilde\bLambda$ and $\tilde\bPsi$ are the maximisers of
$\ell^*(\btheta; \bS)$, the maximisers of
$\ell^*(\dot\btheta; \dot{\bS})$ are $\bL \tilde{\bLambda}$ and
$\bL \tilde{\bPsi} \bL^\top$, respectively. The equivariance under
rotations of the factors is a direct consequence of the invariance of
both $\ell(\btheta; \bS)$ and the penalties in~(\ref{eq:hirose}), when
$\bLambda$ is replaced by $\bLambda \bQ$, for an orthogonal
$q \times q$ matrix $\bQ$.

Despite the above attractive equivariance properties, to our
knowledge, there has been no formal proof that
penalties~(\ref{eq:hirose}) resolve Heywood cases. Furthermore, naive
choice of $\rho$ can introduce considerable finite-sample bias in the
estimation of $\btheta$, as it is illustrated later in the simulations
of Section~\ref{sec:fasimuls}.

\section{Existence of maximum penalised likelihood estimates}
\label{sec:fapl_existence}

Theorem~\ref{thm:existence} provides  general conditions that ensure the
existence of MPL estimates, and use them to examine the properties of
the penalties~(\ref{eq:hirose}).
% We also present conditions, under
% which $\tilde\btheta$ of \eqref{eq:fapl_MPL} has the desirable
% asymptotic properties expected from the ML estimator.

\begin{theorem}[Existence of MPL estimates in factor analysis]
  \label{thm:existence} 
  Let $\bTheta = \left\{\btheta \in \Re^{p (q + 1)}: \theta_m > 0, m > pq \right\}$ and 
  $\partial \bTheta = \{\btheta \in \Re^{p(q + 1)}: \exists m > pq,
  \theta_m = 0 \}$ and
  $\bSigma(\btheta) = \bLambda(\btheta)\bLambda(\btheta)^\top +
  \bPsi(\btheta)$. Assume that $\bS$ has full rank and that the penalty function
  $P^*(\btheta): \bTheta \to \Re$   
  \begin{enumerate}[label=E\arabic*)]
  \item \label{ass:e_cont} is continuous on $\bTheta$;
  \item \label{ass:e_bound} is bounded from above on
    $\bTheta$, i.e. $\underset{\btheta \in \bTheta}{\sup} \, P^*(\btheta)
    < \infty$; and
  \item \label{ass:e_diverge} diverges to $-\infty$ for any sequence
    $\{\btheta(r)\}_{r \in \mathbb{N}}$ such that
    $\lim_{r \to \infty}\btheta(r) \in \partial \bTheta$ and
    $\lim_{r \to \infty}\lambda_{\min}(\bSigma(\btheta(r))) > 0$,
    where $\lambda_{\min}(\bA)$ is the minimum eigenvalue of a matrix $\bA$.
  \end{enumerate}
  Then, the set of MPL estimates
  $\displaystyle \arg \max_{\btheta \in \bTheta}
  \ell^*(\btheta; \bS)$ is non-empty.
\end{theorem}

The proof of Theorem~\ref{thm:existence} is in Section~\ref{appendix:existence} of the supplementary material
document. The theorem is model-agnostic in that it only requires $\bS$
to have full rank and that parameter estimation is conducted using the 
penalised log-likelihood function of \eqref{eq:fapl_MPL}; no assumptions about the true data generating process are made. 
Theorem~\ref{thm:existence} establishes that under
conditions \ref{ass:e_cont}, \ref{ass:e_bound} and \ref{ass:e_diverge}
for the penalty to the log-likelihood, MPL estimation always results
in estimates that are not Heywood cases, in the sense that
$\tilde\btheta$ has $\tilde{\psi}_{j} > 0$ $(j = 1, \ldots, p)$.

The penalties by \citet{akaike:87} and \citet{hirose+etal:2011} in
\eqref{eq:hirose} satisfy assumptions \ref{ass:e_cont},
\ref{ass:e_bound}, and \ref{ass:e_diverge} for $\rho > 0$, and, hence,
MPL estimation using either of those results in no Heywood cases. To
see that, note that matrix inversion, matrix multiplication and trace
are all continuous operations on $\bTheta$. As a result, the penalties
in (\ref{eq:hirose}) are continuous and assumption \ref{ass:e_cont} is
satisfied. The penalties in~\eqref{eq:hirose} can be re-expressed as
\begin{equation}
  \label{eq:hirosereexpr}
  P^*(\btheta) = - \frac{\rho n}{2} \sum_{j = 1}^p \frac{\bA_{jj}(\btheta)}{\bPsi_{jj}(\btheta)} \, ,
\end{equation}
where $\bA_{jj}(\btheta) = \bS_{jj}$ for the \citet{hirose+etal:2011}
penalty, and
$\bA_{jj}(\btheta) = \blambda_j(\btheta)^\top\blambda_j(\btheta)$ for
the \citet{akaike:87} penalty, where $\blambda_j(\btheta)$ is
the $j$th row of $\bLambda(\btheta)$, and $\bC_{jk}$ denotes the
$(j, k)$th element of the matrix $\bC$. Note that
$\bA_{jj}(\btheta) / \bPsi_{jj}(\btheta) \ge 0$ for both
penalties. Hence, (\ref{eq:hirosereexpr}) is bounded above by zero for
$\rho > 0$, and \ref{ass:e_bound} is satisfied.  Now, consider a
sequence $\{\btheta(r)\}_{r \in \mathbb{N}}$ such that
$\lim_{r \to \infty}\btheta(r) \in \partial \bTheta$ and
$\lim_{r \to \infty}\lambda_{\min}(\bSigma(\btheta(r))) > 0$. Then,
there exists at least one $j \in \{1, \ldots, p\}$ such that
$\bPsi_{jj}(\btheta(r)) \to 0$. For $\bA(\btheta) = \bS$, the
penalty~(\ref{eq:hirosereexpr}) diverges to $-\infty$ as
$\bPsi_{jj}(\btheta) \to 0$. For the \citet{akaike:87} version
$\blambda_j(\btheta(r))^\top\blambda_j(\btheta(r))$ can either diverge
to $\infty$ or converge to a constant $c_j > 0$. Only the former can
happen for the chosen sequence $\{\btheta(r)\}_{r \in \mathbb{N}}$,
because, for the latter,
$\blambda_j(\btheta(r))^\top\blambda_j(\btheta(r))$ would need to
converge to zero at an appropriate rate, in which case
$\blambda_j(\btheta(r))^\top\blambda_j(\btheta(r)) +
\bPsi_{jj}(\btheta)$ converges to zero, resulting in
$\bSigma(\btheta(r))$ having at least one zero eigenvalue. Hence,
\ref{ass:e_diverge} is satisfied for both the \citet{akaike:87} and
\citet{hirose+etal:2011} penalties.

Theorem~\ref{thm:fapl_gen_ex} in Section~\ref{appendix:existence} of
the supplementary material document provides an existence result under
more general parameterisations of the factor analysis model, which is
used for proving the consistency results in
Section~\ref{subsec:root_n}, and which might be useful if one wishes
to impose further restrictions on the structure of $\bSigma$, as is
being done, for example, in confirmatory factor analysis (see, for example, \citealt[Chapter~8]{bartholomew.ea:11}).

\section{Asymptotics for maximum penalised likelihood}
\label{sec:fapl_asymp}

% \subsection{Preamble} \label{subsec:preamble} 

% Below we discuss the asymptotic properties of the MPL estimators under appropriate, soft scaling of the penalty function. 
% While we introduced the factor model in Section~\ref{sec:fapl_model} under a normal linear modelling assumption, our results below 
% are more general. 

% In particular, Theorem~\ref{thm:consistency} solely requires that the estimand $\bSigma_0 = \plim_{n \to \infty} \bS$ lies within the model class, i.e. that it admists a representation $\bSigma_0 = \bSigma(\btheta_0)$ for 
% some $\btheta_0 \in \bTheta$. Theorem~\ref{thm:asymptotic_normality}, which operates in restricted model class $\bar{\bTheta}$, we similarly
% only need $\bSigma_0$ to be representable within $\bar{\bTheta}$. 

\subsection{Consistency}
\label{sec:consistency}

To discuss the consistency of estimates for $\bLambda, \bPsi$ in
factor analysis models, we must i) define the estimands
$\bLambda_0, \bPsi_0$ and
$\bSigma_0 = \bLambda_0\bLambda_0^\top + \bPsi_0$ and ii) ensure
identifiability.

If the modelling assumption of Section~\ref{sec:fapl_model} is met for
matrices $\bLambda_0$ and $\bPsi_0$, then the latter are the parameter
values that identify the data generating process. This is the
viewpoint taken in \citet{kano:1983} in their consistency proofs,
where they introduce and use the concept of strong identifiability.

% More
% generally, $\bLambda_0$ and $\bPsi_0$ can be seen as the limits to
% which $\bLambda(\hat\btheta)$ and $\bPsi(\hat\btheta)$, respectively,
% converge in probability as $n\to\infty$, where $\hat\btheta$ is the ML
% estimator of $\btheta$. 

% Forcing identifiability of $\Lambda$ by fixing the rotation if a
% covariance matrix $\bSigma = \bLambda \bLambda^\top + \bPsi$ is close
% to the matrix $\bSigma_0 = \bLambda_0 \bLambda_0^\top + \bPsi_0$, then
% $\bLambda$ and $\bPsi$ are also correspondingly close to $\bLambda_0$
% and $\bPsi_0$. Since we are ultimately interested in consistently
% estimating $\bLambda_0, \bPsi _0$ based on a consistent estimate of
% $\bSigma_0$, such identifiability is of central importance to our
% approach.

Let $\bb{B}$ be any $p \times q$ matrix and $\bb{V} $ be
any positive definite $p \times p$ diagonal matrix and define
$\bSigma = \bb{B} \bb{B}^\top + \bb{V}$. A factor model is said to be
strongly identifiable if and only if, for any $\epsilon > 0$, there
exists a $\delta > 0$ such that
\begin{equation}
  \mnorm{ \bSigma_0 - \bSigma }< \delta \implies \mnorm{\bLambda_0 - \bb{BQ}}< \epsilon \quad \text{and} \quad  \mnorm{\bPsi_0 - \bb{V}} < \epsilon \,,
\end{equation}	
for some orthogonal matrix $\bb{Q}$ of order $q$ and where
$\mnorm{\cdot}$ is some matrix norm. Since our results are derived in a fixed $p$ and fixed $q$ asymptotic regime, 
the particular choice of matrix and vector norm is irrelevant for the developments
(see \citealt[][Corollary~5.4.5]{horn+johnson:2012} and Section~\ref{sec:app_norms} in the supplementary material document for details). 
Hence, throughout, let $\vnorm{\cdot}$ and $\mnorms{\cdot}$ denote any vector and matrix norm, respectively. 

\begin{theorem}
  \label{thm:consistency}

  Assume that 
  \begin{enumerate}[label = C\arabic{*})]
  \item \label{ass:si}
    the factor model is strongly identifiable;
  \item \label{ass:exist_cons} the set of maximum penalised likelihood
    estimates
    $\arg \max_{\btheta \in \bTheta} \ell^*(\btheta; \bS) $ is non-empty; and
  \item \label{ass:nonpos} $P^*(\btheta) \leq 0$ for all
    $\btheta \in \bTheta$.
  \end{enumerate}
  Then, for any $\epsilon > 0$, there exists a $\delta>0$ such that
  \begin{equation}
    \mnorm{\bS - \bSigma_0} < \delta \quad \text{and} \quad |n^{-1}  P^*(\btheta_0)|< \delta \implies \mnorm{\bLambda_0 -\bLambda(\bttilde)  \bb{Q} } < \epsilon \quad \text{and} \quad \mnorm{\bPsi_0 - \bPsi(\bttilde)} < \epsilon \,,
  \end{equation}   
  for some orthogonal $q \times q$ matrix $\bb{Q}$.
\end{theorem}

The proof of Theorem~\ref{thm:consistency} is in
Section~\ref{sec:app_cons} of the supplementary material document.
Theorem~\ref{thm:consistency} shows that if $\bS \to \bSigma_0$ and
$n^{-1} P^*(\btheta_0) \to 0$ either in probability or almost surely,
then the MPL estimates $\bLambda(\tilde\btheta)$,
$\bPsi(\tilde\btheta)$ converge to $\bLambda_0$, $\bPsi_0$,
respectively, in probability or almost surely, up to orthogonal
rotations of $\bLambda(\tilde\btheta)$. Note that the conditions that
we require of the penalty function are mild; $P^*(\btheta)$ can be
deterministic or depend on the responses, as long as it is pointwise
$o_p(n)$.

\subsection{$\sqrt{n}$-consistency and asymptotic distribution}
\label{subsec:root_n}

Results on the rate of consistency and the asymptotic distribution of
the MPL estimator can be derived under a stronger condition on the
order of the penalty than that of Theorem~\ref{thm:consistency}. 
In particular, we are interested in $\sqrt{n}$-consistency of the MPL estimates, i.e. 
$$
  \sqrt{n}\mnorms{\bLambda(\bttilde)\bQ_1 - \bLambda(\bthat)\bQ_2} = o_p(1), \quad \sqrt{n}\mnorms{\bPsi(\bttilde) - \bPsi(\bthat)} = o_p(1)\,,
$$ 
where $\bttilde$ and $\bthat$ denote the MPL and ML 
estimates, respectively, and $\bQ_1, \bQ_2$ are appropriate sequences of orthogonal rotation matrices. Under stronger identification conditions, we can also establish results on the asymptotic distribution of $\bttilde$ and $\bthat$. 

Central to such results is the local identification restriction that the Jacobian of $\vect(\bSigma(\btheta))$ has full column rank at the parameter of interest $\btnod$.
If the map $\btheta \mapsto \vect(\bSigma(\btheta))$ is continuously differentiable with full column rank Jacobian at $\btnod$, then there exists an open neighbourhood $\mathcal{U}_0$ around $\btnod$, 
such that for any $\varepsilon > 0$, there exists a $\delta > 0$, with
$$
    \mnorm{\bSigma(\btheta) - \bSigma(\btnod)} < \delta \implies \vnorm{\btheta - \btnod} < \varepsilon \,, 
$$
for all $\btheta \in \mathcal{U}_0$.  This assumption ensures that the
information matrix is invertible, which is required for
$\sqrt{n}$-asymptotics.  The full column rank condition~\ref{ass:si_2}
in Theorem~\ref{thm:asymptotic_normality} is, for example, also
present in \citet[][Theorem~2 and Theorem~3]{anderson+amemiya:1988},
who establish the asymptotic normal distribution of the
$\sqrt{n} (\hat\btheta - \btheta_0)$ in factor analysis models under
that and additional conditions.

In the unrestricted model where $\bTheta = \{\btheta \in \Re^{p(q + 1)}: \theta_m > 0, m > pq\}$, 
this full rank condition does not hold due to the invariance of the variance-covariance matrix under orthogonal rotations of $\bLambda$. 
% When $q\geq 2$, this invariance is continuous: there exist smooth one-parameter families $\bQ(t)$ with $\bQ(0)=\bI_q$ such that 
% for $\btheta_t = (\vect(\bLambda \bQ(t))^\top, \diag (\bPsi)^\top)^\top$, $\bSigma(\btheta_t)$ is constant in $t$, 
% and hence the Jacobian of $\btheta \mapsto \vect(\bSigma(\btheta_t))$ cannot have full column rank at $\btheta_t$ \IK{I am a bit lost; we one-parameter?}. 
Thus, henceforth, we focus on a restricted parameter space  $\bar \bTheta \subseteq \Re^d$ with covariance mapping $\bSigma(\btheta) = \bLambda(\btheta) \bLambda(\btheta)^\top + \bPsi(\btheta)$ that does not allow for a rank-deficient Jacobian. We further assume that $\bar\bTheta$ is contained in $\bTheta$ in the sense that for every $\btheta \in \bTheta$, there exists a $\bar\btheta \in \bar\bTheta$ and an orthogonal rotation $\bb Q$, such that 
$\bLambda(\btheta)\bQ = \bLambda(\bar \btheta)$ and $\bPsi(\btheta) = \bPsi(\bar \btheta)$. Common parameter restrictions encountered in practice, 
that accommodate this requirement include requiring $\bLambda$ to be upper-triangular, or requiring $\bLambda^\top \bPsi^{-1} \bLambda$ to be diagonal \citep[see, for example,][Chapter~3]{bartholomew.ea:11}. 
% Such restrictions remove the continuous rotation degrees of freedom, though discrete indeterminacies, like sign flips and permutations, may remain. 
% These are, however, compatible with a full-rank Jacobian.

Note that the parameter vector defining the matrices $\bLambda_0$ and $\bPsi_0$ need not itself be in $\bar\bTheta$. Rather, we assume that $\bar\bTheta$ contains a parameter vector that corresponds to a loading matrix that can be appropriately rotated to give $\bLambda_0$. Unless $\bLambda_0$ and $\bPsi_0$ correspond to a parameter vector in $\bar\bTheta$, convergence of loading matrices  is therefore naturally stated after appropriate rotation.

\begin{theorem}
	\label{thm:asymptotic_normality}
	Assume that
	\begin{enumerate}[label=N\arabic*)]
		\item \label{ass:S_con} there exists a $\btheta_0 \in \bar \bTheta$ such that
		$\bS \overset{p}{\longrightarrow} \bSigma(\btheta_0)$ as $n \to \infty$;
		\item \label{ass:si_2} $\bSigma(\btnod)$ is strongly identifiable in $\bar \bTheta$ and the Jacobian of $\textrm{vec}(\bSigma(\btheta))$ with respect to $\btheta$ has full column rank at $\btnod$; 
		\item the set of maximum penalised likelihood estimates  $\arg \underset{\btheta \in \bar \bTheta}{\max} \, \{ \ell^*(\btheta; \bS) \}$ is not empty; and
		\item \label{ass:pen_n} $P^*(\btheta) = c_n P(\btheta)$ where $P(\btheta)$ is nonpositive, deterministic, invariant under orthogonal rotations of $\bLambda$ and
			continuously differentiable on $\bar\bTheta$, with $c_n = o_p(\sqrt{n})$ positive.
	\end{enumerate}
	Then, there exist sequences of orthogonal rotation matrices $\bQ_1, \bQ_2$ such that: 
	\begin{equation}
		\mnorm{\bLambda(\bttilde)\bQ_1 - \bLambda_0} \overset{p}{\longrightarrow} 0, \quad \mnorm{\bPsi(\bttilde)  - \bPsi_0} \overset{p}{\longrightarrow} 0 \,,
	\end{equation}
	and 
	\begin{equation}
		\sqrt{n}\mnorm{\bLambda(\bttilde)\bQ_1 -\bLambda(\bthat)\bQ_2} \overset{p}{\longrightarrow} 0, \quad \sqrt{n}\mnorm{\bPsi(\bttilde)  - \bPsi(\bthat)} \overset{p}{\longrightarrow} 0 \,,
	\end{equation}
  where $\bthat$ and $\bttilde$ denote two sequences of the ML and MPL estimates, respectively. 
\end{theorem}

The proof of Theorem~\ref{thm:asymptotic_normality} is in
Section~\ref{subsec:r_n_cons} of the supplementary material document. Theorem~\ref{thm:asymptotic_normality} as presented above gives conditions about the consistency of ML and MPL estimators, as well as
the rate of convergence to one another. If one is interested in establishing the asymptotic distribution of the MPL estimator $\bttilde$ based on the limiting distribution of the 
ML estimator $\bthat$, one can replace the strong identifiability condition on $\bLambda_0, \bPsi_0$ with a more stringent point-identification condition on 
$\btheta_0$: 
\begin{enumerate}
  \item[N2$^*$)] for any $\epsilon>0$, there exists a $\delta>0$ such that for all $\btheta \in \bar\bTheta$, $\mnorms{\bSigma(\btheta) - \bSigma(\btheta_0)} < \delta$ 
implies that $\vnorm{\btheta - \btnod} < \epsilon$ and the Jacobian of $\textrm{vec}(\bSigma(\btheta))$ with respect to $\btheta$ has full column rank at $\btnod$; 
\end{enumerate}
In this instance the conclusion of Theorem~\ref{thm:asymptotic_normality} is that
$\bttilde \overset{p}{\longrightarrow} \btnod$ and $\sqrt{n}\vnorm{\bttilde - \bthat} = o_p(1)$.
This stronger identification 
condition may be required when one wishes to establish the asymptotic distribution of $\bttilde, \bthat$.
Slutsky's lemma
then implies that if $\sqrt{n} (\hat\btheta - \btheta_0)$ has a normal
distribution asymptotically, then
$\sqrt{n} (\tilde\btheta - \btheta_0)$ with a penalty scaled as in
\ref{ass:pen_n} has the same asymptotic distribution.

Note that the conditions of Theorem~\ref{thm:consistency} and
Theorem~\ref{thm:asymptotic_normality} only require that the limit of
the estimator of the variance-covariance matrix $\bS$ decomposes in a
exploratory factor analysis fashion, and normality of errors is not
essential.

\section{Maximum softly penalised likelihood}
\label{sec:mspl}

\begin{table}[t]
  \centering 
  \caption{Loading matrix settings $A_3$ and $B_3$.} 
  \label{tab:loadings}
  \begin{small}
  \begin{tabular}{CCCCCCCC}
    \toprule
    \text{Item} & \multicolumn{3}{c}{\text{Setting $A_3$}} & & \multicolumn{3}{c}{\text{Setting $B_3$}} \\ 
    \cmidrule{2-4} \cmidrule{6-8}
    1 & 0.80 & 0  & 0   && 0.80 & 0  & 0 \\
    2 & 0.65 & 0  & 0   && 0.80 & 0  & 0 \\
    3 & 0.45 & 0  & 0   && 0.80 & 0  & 0 \\
    4 & 0  & 0.80 & 0   && 0  & 0.80 & 0 \\
    5 & 0  & 0.65 & 0   && 0  & 0.80 & 0 \\
    6 & 0  & 0.45 & 0   && 0  & 0.80 & 0 \\
    7 & 0  & 0  & 0.80  && 0  & 0  & 0.30 \\
    8 & 0  & 0  & 0.65  && 0  & 0  & 0.30 \\
    9 & 0  & 0  & 0.45  && 0  & 0  & 0.30 \\ \bottomrule
  \end{tabular}
\end{small}
\end{table}

Theorem~\ref{thm:existence} establishes conditions on $P^*(\btheta)$
that ensure the existence of the MPL estimates. On the other hand,
Theorem~\ref{thm:consistency} and
Theorem~\ref{thm:asymptotic_normality} involve sufficient conditions
on the order of the penalty $P^*(\btheta)$ for the consistency of the
MPL estimator. Specifically, if $P^*(\btheta) = o_p(\sqrt{n})$ the
respective order conditions in Theorem~\ref{thm:consistency} and
Theorem~\ref{thm:asymptotic_normality} are satisfied.

Suppose that $P^*(\btheta) = c_n P(\btheta)$, where the functional
part $P(\btheta)$ satisfies the conditions of
Theorem~\ref{thm:existence} for the existence of MPL estimates, and
where $c_n > 0$ is a scaling factor. One way to derive a suitable
scaling factor is to consider how information about the model
parameters accumulates as $n$ increases.  For example, we can derive a
principled heuristic for $c_n$ by considering the exploratory factor
analysis model in~\eqref{eq:loglikl2} under independence
(i.e.~$\bLambda$ is a matrix of zeros). The unknown parameters are the
vector of variances $\sigma_1^2, \ldots \sigma_p^2$, and the
information matrix about those parameters is a diagonal matrix with
$j$th diagonal element $n / (2 \sigma^4_j)$. Standard results on the
asymptotic distribution of the ML estimator give that
$\sqrt{n / 2} (s_j^2 / \sigma_j^2 - 1)$ converges in distribution to a
standard normal random variable for all $j \in \{1, \ldots, p\}$. The
rate of information accumulation for each coordinate is $\sqrt{n / 2}$
and, hence, we can choose $c_n = \sqrt{2 / n}$.  This choice is in
line with the requirements of Theorem~\ref{thm:consistency} and
Theorem~\ref{thm:asymptotic_normality} for any exploratory factor
analysis model, while ensuring that the penalisation strength is
asymptotically negligible. We call maximum softly-penalised likelihood (MSPL)
estimation, MPL estimation with asymptotically negligible penalties
that guarantee existence and $\sqrt{n}$-consistency.  In
Section~\ref{sec:fapl_existence} we showed that the conditions for the
existence of the MPL estimates are satisfied for the
penalties~(\ref{eq:hirose}) in \citet{akaike:87} and
\citet{hirose+etal:2011}. In particular, both penalties have the form
$P^*(\btheta) = \rho n P(\btheta) / 2$, $\rho > 0$, $P(\btheta) \le 0$
and $P(\btheta) = O(1)$. Both penalties can be adapted for MSPL
estimation by setting $\rho = 2 \sqrt{2 / n^{3}}$.

\section{Simulation studies} \label{sec:fasimuls}

\begin{figure}[t]
  \begin{center}
    \includegraphics[width=\textwidth]{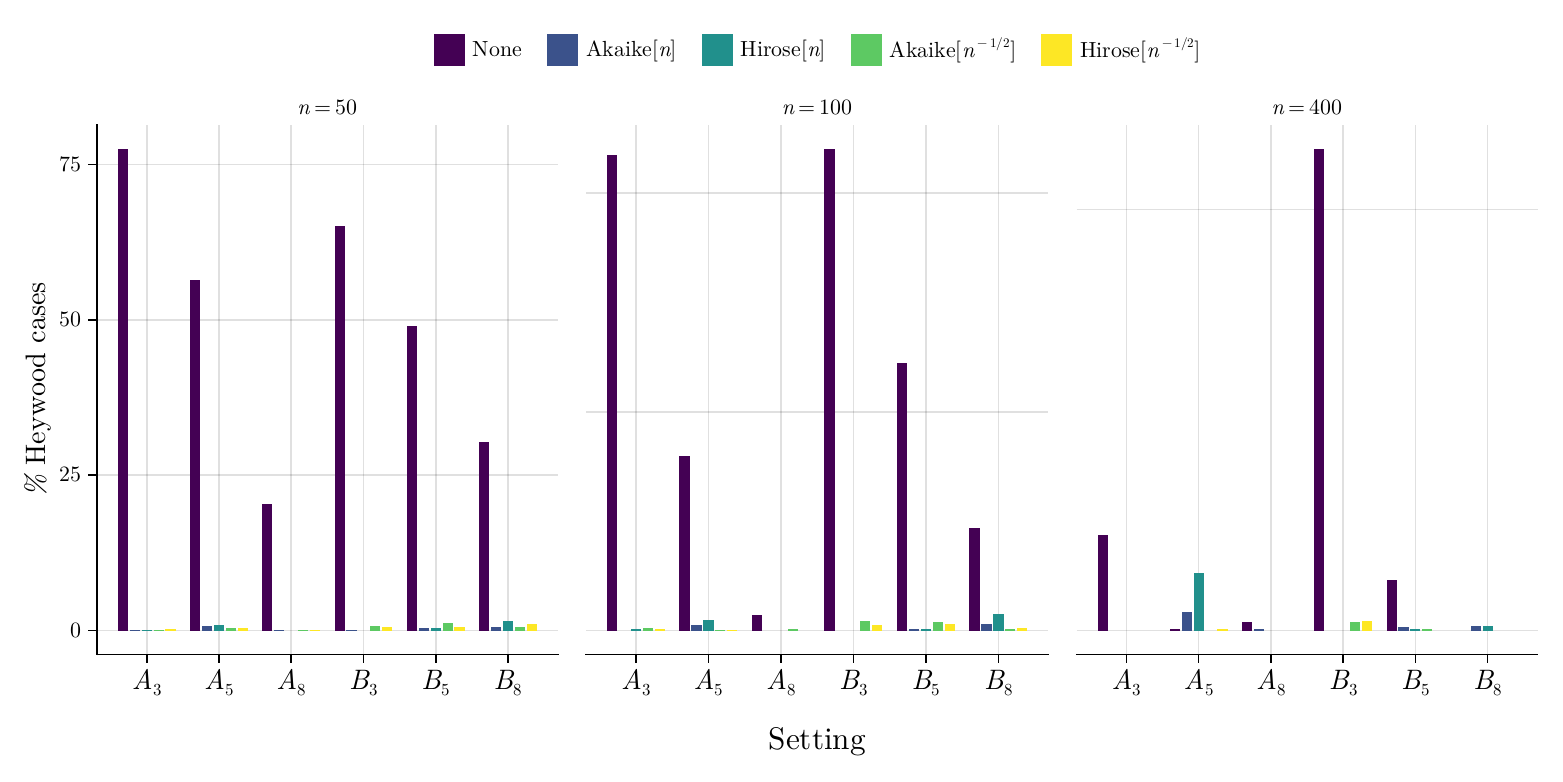}
  \end{center}
  \caption{Percentage of samples (out of $1000$) that have been
    identified as Heywood cases for ML (``None''), MPL with
    Akaike[$n$] and Hirose[$n$] penalties, and MSPL with
    Akaike[$n^{-1/2}$] and Hirose[$n^{-1/2}$] penalties,
    $n \in \left\{50,100,400\right\}$, and loading matrix settings
    $A_3$, $B_3$, $A_5$, $B_5$, $A_8$, and $B_8$.}
  \label{fig:perc_used}
\end{figure}

We conduct a series of simulation experiments to compare the
performance of ML estimation, MPL estimation based on the
penalties~(\ref{eq:hirose}) with vanilla choices of $\rho$, and MSPL
estimation. The methods are compared in terms of their ability to
handle Heywood cases, frequentist properties of the estimators, and
selection of the number of factors when Heywood cases are present.

Our simulation settings have been informed by the simulation-based
results reported in \citet{cooperman.waller:22}. Specifically,
\citet{cooperman.waller:22} identified, through an extensive
simulation study, the following causes of Heywood cases in order of
importance: item-to-factor ratio, model specification (correct,
fitting one factor less than the actual number, and fitting one extra
factor than the actual number), sample size and loading matrix pattern
(low against high loadings on the same factor, and factors with all low
loadings against factors with all high loadings).

We fix the number of factors to $q = 3$. We then consider sample sizes
$n \in \left\{50, 100, 400 \right\}$ and the item-to-factor ratios
$3:1$, $5:1$, and $8:1$, and let the loading matrix $\bLambda$ vary
across experimental settings. Specifically, for the loading matrix
with item-to-factor ratio $3:1$, we use the matrices in settings $A_3$
and $B_3$ in Table~\ref{tab:loadings}, which are motivated from the
settings of \citet[Table~2]{cooperman.waller:22}. Setting $A_3$
decreases the factor loadings sequentially, while setting $B_3$
assumes two strong and one weak factor. Setting $A_5$ for the $5:1$
ratio and setting $A_8$ for the $8:1$ ratio are defined
correspondingly to $A_3$, where we choose the nonzero column blocks to
be $(0.80, 0.65, 0.50, 0.35, 0.20)$ and
$(0.80, 0.70, 0.60, 0.50, 0.40, 0.30, 0.20, 0.10)$, respectively. For
settings $B_5$ for the $5:1$ ratio and $B_8$ for the $8:1$ ratio, we
simply repeat the non-zero loadings in Table~\ref{tab:loadings}
according to the item-to-factor ratio. The specific variances $\bPsi$
are set so that the diagonal elements of
$\bSigma = \bLambda\bLambda^\top + \bPsi$ are all one.

We compare the ML estimator with the MSPL estimators using
appropriately scaled versions of the penalties~(\ref{eq:hirose}) with
$\rho = 2\sqrt{2 / n^{3}}$ based on the discussion in
Section~\ref{sec:mspl}. We refer to those penalties as
``Akaike[$n^{-1/2}$]'' and ``Hirose[$n^{-1/2}$]''.  We also
consider MPL with the non-decaying scaling $\rho = 1$, which was also
used in \citet{hirose+etal:2011}. We refer to those penalties as
``Akaike[$n$]'' and ``Hirose[$n$]''.

For each combination of loading matrix, sample size, and
item-to-factor ratio, we draw $1000$ independent samples according to
the factor analysis model~(\ref{eq:fa_model}) and, for each sample, we
compute the estimates of \eqref{eq:fapl_MPL}. The estimates are
computed by first getting MPL estimates from $100$ iterations of an
EM-maximisation of the penalised log-likelihood, which we then use as
starting values for a Newton-Raphson optimisation routine. We identify
Heywood cases heuristically, when at least one of the following
occurs: the estimation procedure fails, the normalised gradient
$\{\nabla\nabla^\top\ell^*(\btheta)\}^{-1} \nabla\ell^*(\btheta)$ has
at least one element with absolute value greater than $10^{-4}$, and at
least one of the estimates of $\psi_{1}, \ldots, \psi_{p}$ is less
than $10^{-4}$.

\begin{figure}[t]
  \centering 
   \begin{center}
  	\includegraphics[width=\textwidth]{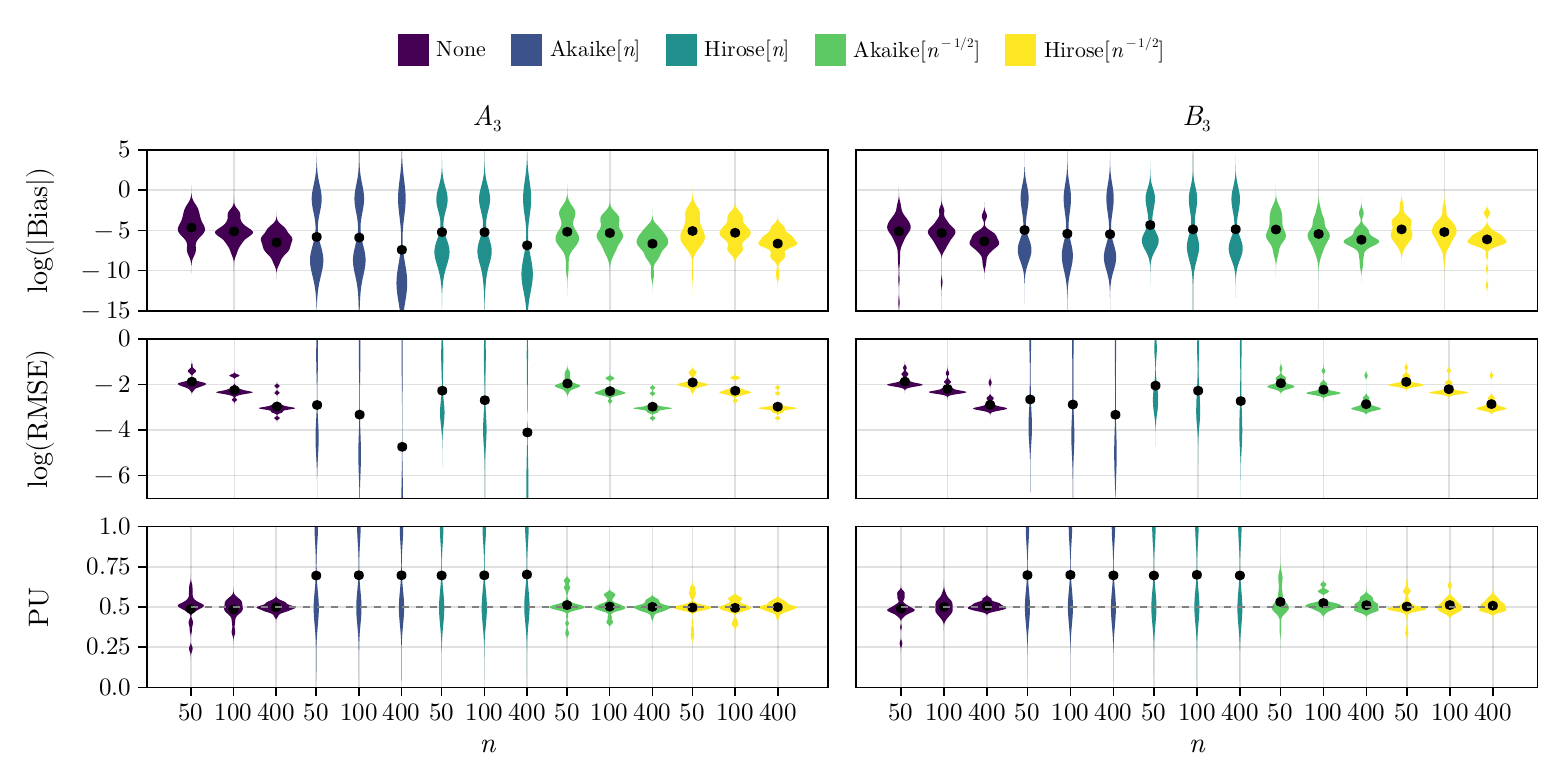}
  \end{center}
  \caption{Violin plots of estimates of
    $\log(|\mathrm{Bias}|)$ (top panel), $\log(\mathrm{RMSE})$ (middle
    panel) and probability of underestimation (bottom panel) for the
    elements of $\bLambda\bLambda^\top$, for each estimator,
    $n \in \left\{50,100,400\right\}$, and loading matrix settings
    $A_3$ and $B_3$. The average over all elements for each setting is
    noted with a dot.}
  \label{fig:combined}
\end{figure}

Figure~\ref{fig:perc_used} shows the percentage of samples that have
been identified as Heywood cases. It is evident that ML estimation
results in a considerable number of Heywood cases across experimental
settings, while, as expected from Theorem~\ref{thm:existence}, MPL and
MSPL estimation are effective in dealing with them. The negligible
fraction of estimates that are identified as Heywood cases for MPL and
MSPL estimation are attributable to the heuristics we use for the
identification of heywood cases, and can be eliminated by less
stringent heuristics.

We evaluate the finite-sample performance of ML, MPL and MSPL
estimators in terms of bias, probability of underestimation and root
mean-squared error (RMSE), estimated excluding the samples that have
been identified as Heywood cases.

Figure~\ref{fig:combined} shows violin plots of coordinatewise
estimates of the logarithm of the absolute bias, the logarithm of the
RMSE, and the probability of underestimation of the unique elements of
$\bLambda\bLambda^\top$, for each estimator,
$n \in \left\{50, 100, 400\right\}$, and loading matrix settings $A_3$
and $B_3$. A black dot indicates the average over all coordinates in
each specification. The MSPL estimates, with soft scaling of order
$n^{-1/2}$ exhibit the smallest or close to the smallest bias across
all methods and a rate of decay that is in line with
Theorem~\ref{thm:asymptotic_normality}. Notably, the Akaike[$n$] and
Hirose[$n$] penalised estimators exhibit large finite sample bias. We
also see that the MSPL estimators are well calibrated, with a
probability of underestimation close to $1/2$ across all settings. In
contrast, the Akaike[$n$] and Hirose[$n$] estimators consistently
underestimate the elements of $\bLambda \bLambda^\top$. This
underestimation is expected from the excessive penalisation that
results from using a scaling factor of order $n$. Similarly to bias,
the RMSE of the MSPL estimates is the lowest or close to the minimal RMSE
across all methods. The coordinatewise estimates of the logarithm of
absolute bias, the logarithm of RMSE, and probability of
underestimation for loading matrix settings $A_5$ and $B_5$, and $A_8$
and $B_8$ are shown in Figure~\ref{fig:combined2} and
Figure~\ref{fig:combined3} of the supplementary material document,
respectively. The findings are the same as above for $A_3$ and $B_3$.

\begin{figure}[t]
  \begin{center}
\includegraphics[width=\textwidth]{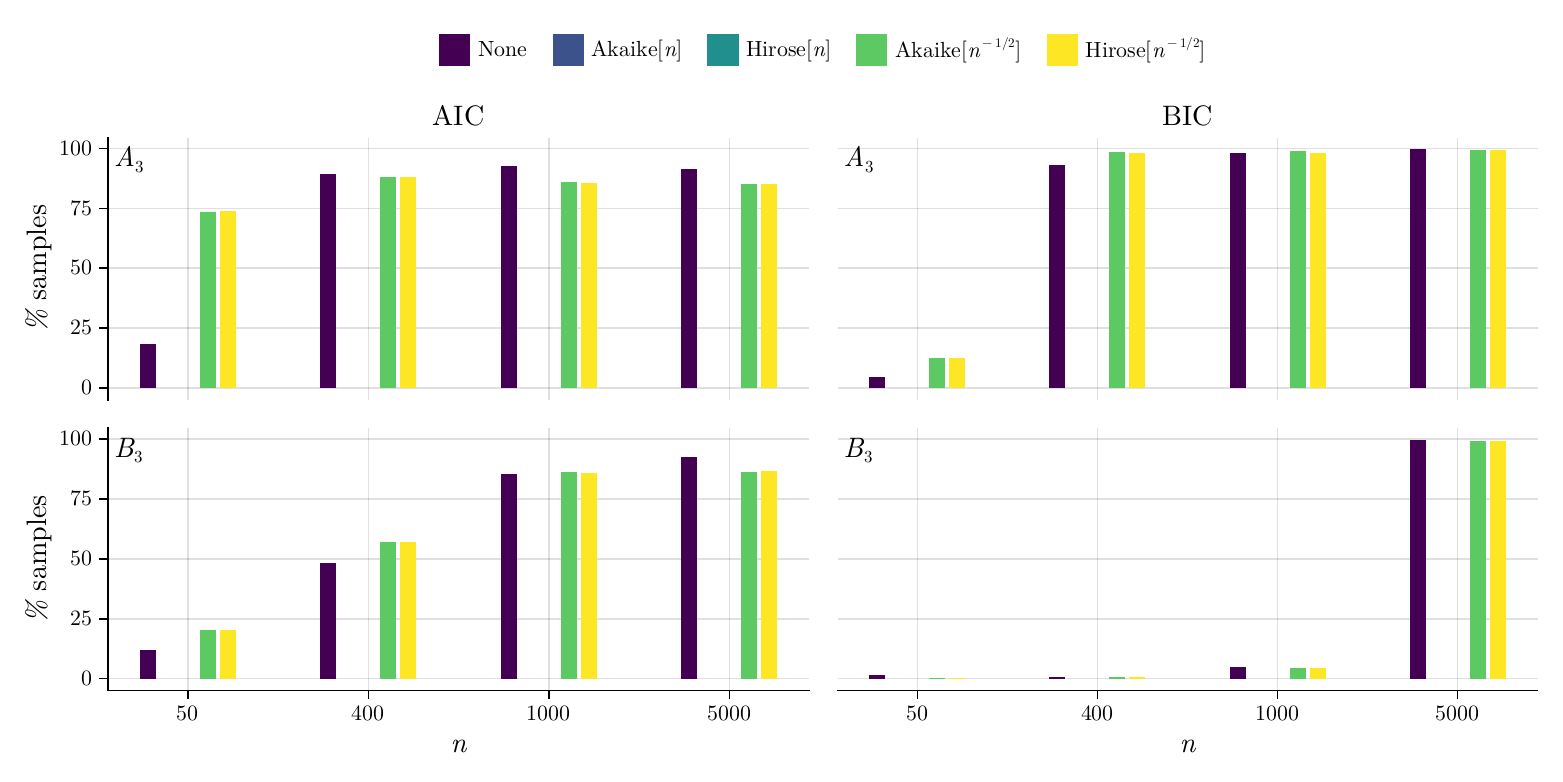}
  \end{center}
  \caption{Percentage of times the model with $3$ factors is
    selected for each estimator, $n \in \{50, 400, 1000, 5000 \}$,
    and loading matrix settings $A_3$ and $B_3$, using AIC and BIC. The absence of 
    vertical bars pertaining to the Akaike[$n$] and Hirose[$n$] based model selection procedures
    indicate that these methods have never selected the correct model.}
  \label{fig:ms}
\end{figure}

We also assess the performance of AIC and BIC selection of the number
of factors based on each estimator (see \citealt{akaike:87} and
\citealt{hirose+etal:2011} for details on these criteria for factor
models) for item-to-factor ratio $3:1$ with settings $A_3$ and $B_3$
for the loading matrix, and $n \in \{50, 400, 1000, 5000\}$.  We note
that when fitting models with $q \in \{1,2\}$, the model is
misspecified and the assumptions of Theorem~\ref{thm:consistency} and
Theorem~\ref{thm:asymptotic_normality} do not apply.  This design
choice mirrors common practitioner behaviour in factor selection and
motivates our inclusion of this scenario.  We fit the factor analysis
model~\eqref{eq:fa_model} for $q \in \{1, \ldots, 5\}$. The AIC and
BIC are computed from the unpenalised log-likelihood evaluated at the
ML and MSPL estimates, respectively.  We use all the samples,
including those that have been identified as leading to Heywood cases,
for computing AIC and BIC. Figure~\ref{fig:ms} shows the percentage of
times that the model with $3$ factors was selected with AIC and BIC at
each estimator and $n \in \left\{50, 400, 1000, 5000\right\}$. We note
that AIC- and BIC-based model selection performs as expected with the
MSPL estimators with Akaike[$n^{-1/2}$] and Hirose[$n^{-1/2}$]
penalties. BIC-based model selection selects the correct model with
increasing probability as $n$ increases, which is the result of the
consistency of BIC-based model selection \citep[see, for
example][Chapter~4]{claeskens+hjort:2008}. For $n = 50$, model
selection based on MSPL estimators is also found to outperform that
based on the ML estimator, most probably due to the strong handling of
Heywood cases, even in small samples. On the other hand, the MPL
estimators with Akaike[$n$] and Hirose[$n$] penalties result in poor
performance in both AIC- and BIC-based model selection, mainly due to
the strength of the penalty.

For $n < 5000$, we observe poor performance of BIC in selecting the correct number of factors
in setting $B_3$ across all estimators which we attribute to 
the two strong and one extremely weak factor (see Table~\ref{tab:loadings}). As
Table~\ref{tab:ms_qbic} shows, BIC identifies the two strong factors
in the majority of cases for small $n$, and starts identifying the
weak factor more frequently slowly as $n$ increases with ML or
MSPL estimators.

\begin{table}[t]
  \centering
  \caption{Percentage of times each number of factors has been
    selected using minimum BIC, for ML and MSPL with Akaike[$n$], Hirose[$n$],
    Akaike[$n^{-1/2}$] and Hirose[$n^{-1/2}$] penalties, under loading
    matrix setting $B_3$ and $n \in \left\{50,400, 1000, 5000\right\}$.}
  \label{tab:ms_qbic}
  \begin{small}
  \begin{tabular}{CCRRRRR}
    \toprule
    n & q & \text{None} &  \text{Akaike$[n]$} & \text{Hirose$[n]$} & \text{Akaike$[n^{-1/2}]$} & \text{Hirose$[n^{-1/2}]$} \\ 
    \midrule
    50 & 1 & 5.8 & 100.0 & 100.0 & 1.5 & 1.5\\
    & 2 & 92.6 & 0.0 & 0.0 & 98.2 & 98.2\\
    & 3 & 1.5 & 0.0 & 0.0 & 0.3 & 0.3\\
    & 4 & 0.1 & 0.0 & 0.0 & 0.0 & 0.0\\ \midrule 
    400 & 1 & 0.0 & 4.7 & 65.1 & 0.0 & 0.0\\
    & 2 & 99.5 & 95.3 & 34.9 & 99.5 & 99.5\\
    & 3 & 0.5 & 0.0 & 0.0 & 0.5 & 0.5\\
    & 4 & 0.0 & 0.0 & 0.0 & 0.0 & 0.0\\ \midrule 
    1000 & 1 & 0.0 & 0.0 & 0.0 & 0.0 & 0.0\\
    & 2 & 95.2 & 100.0 & 100.0 & 95.3 & 95.4\\
    & 3 & 4.8 & 0.0 & 0.0 & 4.7 & 4.6\\
    & 4 & 0.0 & 0.0 & 0.0 & 0.0 & 0.0\\ \midrule 
    5000 & 1 & 0.0 & 0.0 & 0.0 & 0.0 & 0.0\\
    & 2 & 0.0 & 100.0 & 100.0 & 0.0 & 0.0\\
    & 3 & 100.0 & 0.0 & 0.0 & 100.0 & 100.0\\
    & 4 & 0.0 & 0.0 & 0.0 & 0.0 & 0.0\\
    \bottomrule
  \end{tabular}
\end{small}
\end{table}

\section{Real data examples}
\label{sec:fapl_examples}

We estimate the factor model~(\ref{eq:fa_model}) using ML, and MSPL
using the Akaike[$n^{-1/2}$] and Hirose[$n^{-1/2}$] penalties for
three data sets where Heywood cases have been encountered in published
work. The data sets are i) the Davis data \citep{rao:55}, which
involves $n = 421$ observations and $p = 9$ items, ii) the Emmett data
\citep{emmett:49, lawley.maxwell:71} which involves $n = 211$
observations and $p = 9$ items, and iii) the Maxwell data
(\citealt{maxwell:61}, and \citealt[][p. 44]{lawley.maxwell:71}),
which involves $n = 810$ observations and $p = 10$ items. Heywood
cases result in the ML estimates of the factor
model~\eqref{eq:fa_model} with $q = 2$ for the Davis data, $q = 5$ for
the Emmett data, and $q = 4$ for the Maxwell data. The three data sets
have been also analysed in \citet{akaike:87}.

Table~\ref{tab:estimates} gives the estimates of the communalities
$\sum_{k = 1}^q \lambda_{jk}^2$ $(j = 1, \ldots, p)$ using ML, and
MSPL with Akaike$[n^{-1/2}]$ and Hirose$[n^{-1/2}]$ penalties, across
different number of factors, along with the corresponding AIC and BIC
values. As expected, ML estimation can lead to Heywood cases, which
manifest as atypically large estimated communalities. In contrast, and
as expected, there are no Heywood cases when MSPL estimation is used,
and communality estimates are reasonable with no substantial impact on
AIC and BIC values. Specifically, in the Davis data set, item $1$ has
an atypically large communality ML estimate for $q = 2$,
while the MSPL estimates are all within reasonable ranges. MSPL
estimation also resolves the Heywood cases that result in atypically
large ML communality estimates for the Emmett data for $q = 4$ and
$q = 5$, and the Maxwell data for $q =
4$. Table~\ref{davisdataresults_full},
Table~\ref{emmettdataresults_full} and
Table~\ref{maxwelldataresults_full} in the supplementary material
document show the estimates of $\bPsi$ and $\bLambda$ for the Emmett
data, for $q \in \{1, \ldots, 5\}$. As is apparent, the large ML
estimated communalities for $q = 4$ and item 3, and $q = 5$ and item
4, correspond to negative ML variance estimates for those items. In
contrast, and as expected all MSPL variance estimates are
positive. Notably, due to soft penalisation, the ML and MSPL estimates
that do not correspond to Heywood cases are similar.

\begin{table}[H]
  \centering
  \caption{Estimated communalities ($\times 10^3$) for the Davis, Emmett, and Maxwell data, using ML, and MSPL with Akaike[$n^{-1/2}$] and Hirose[$n^{-1/2}$] penalties, for $q \in \{1, \ldots, 5\}$, with AIC and BIC values. Heywood cases are shown in bold.}
  \label{tab:estimates}
  \begin{small}
  \begin{tabular}{CCRRRRRRRRRRRR}
					\toprule
					&  & \multicolumn{10}{c}{Item} &  \\ \cmidrule{3-12}
					\textrm{Method} & q & \multicolumn{1}{c}{1} & \multicolumn{1}{c}{2} & \multicolumn{1}{c}{3} & \multicolumn{1}{c}{4} & \multicolumn{1}{c}{5} & \multicolumn{1}{c}{6} & \multicolumn{1}{c}{7} & \multicolumn{1}{c}{8} & \multicolumn{1}{c}{9} & \multicolumn{1}{c}{10} & \textrm{AIC} & \textrm{BIC} \\ \midrule
					\multicolumn{14}{c}{\textrm{Davis data}} \\ \midrule
					\text{ML} & 1 & 658 & 661 & 228 & 168 & 454 & 800 & 705 & 434 & 703 &  & 1694.69 & 1767.46 \\
					& 2 & \textbf{14145} & 634 & 224 & 176 & 463 & 813 & 705 & 439 & 702 &  & 1683.02 & 1788.12 \\ \midrule
					\text{Akaike[$n^{-1/2}$]} & 1 & 654 & 657 & 227 & 167 & 452 & 795 & 701 & 432 & 699 &  & 1694.70 & 1767.47 \\
					& 2 & 877 & 660 & 226 & 185 & 467 & 808 & 698 & 438 & 696 &  & 1684.89 & 1790.00 \\ \midrule
					\text{Hirose[$n^{-1/2}$]} & 1 & 658 & 661 & 228 & 168 & 454 & 799 & 705 & 434 & 703 &  & 1694.69 & 1767.46 \\
					& 2 & 882 & 664 & 227 & 187 & 470 & 813 & 703 & 441 & 701 &  & 1684.88 & 1789.99 \\
					\midrule
					\multicolumn{14}{c}{\textrm{Emmett data}} \\ \midrule
					\text{ML} & 1 & 510 & 537 & 300 & 548 & 390 & 481 & 525 & 224 & 665 &  & 1176.56 & 1236.90 \\
					& 2 & 538 & 536 & 332 & 809 & 592 & 778 & 597 & 256 & 782 &  & 984.56 & 1071.71 \\
					& 3 & 550 & 573 & 383 & 788 & 619 & 823 & 600 & 538 & 769 &  & 977.58 & 1088.20 \\
					& 4 & 544 & 556 & \textbf{14307} & 797 & 612 & 800 & 604 & 737 & 773 &  & 986.51 & 1117.23 \\
					& 5 & 547 & 645 & 376 & \textbf{8877} & 553 & 991 & 681 & 532 & 759 &  & 992.27 & 1139.76 \\ \midrule
					\text{Akaike[$n^{-1/2}$]} & 1 & 505 & 531 & 297 & 543 & 386 & 476 & 519 & 222 & 658 &  & 1176.57 & 1236.91 \\
					& 2 & 532 & 528 & 329 & 797 & 585 & 768 & 590 & 253 & 770 &  & 984.59 & 1071.74 \\
					& 3 & 543 & 565 & 379 & 777 & 612 & 809 & 593 & 532 & 758 &  & 977.62 & 1088.24 \\
					& 4 & 540 & 557 & 600 & 789 & 611 & 797 & 597 & 615 & 763 &  & 986.85 & 1117.58 \\
					& 5 & 547 & 632 & 405 & 820 & 653 & 824 & 646 & 492 & 758 &  & 992.74 & 1140.23 \\ \midrule
					\text{Hirose[$n^{-1/2}$]} & 1 & 510 & 537 & 300 & 548 & 390 & 481 & 524 & 224 & 665 &  & 1176.57 & 1236.90 \\
					& 2 & 538 & 536 & 333 & 807 & 592 & 778 & 597 & 256 & 780 &  & 984.57 & 1071.72 \\
					& 3 & 550 & 573 & 384 & 788 & 620 & 820 & 600 & 536 & 768 &  & 977.60 & 1088.22 \\
					& 4 & 547 & 565 & 605 & 800 & 620 & 807 & 605 & 619 & 773 &  & 986.83 & 1117.56 \\
					& 5 & 555 & 641 & 410 & 833 & 662 & 836 & 654 & 496 & 769 &  & 992.72 & 1140.20 \\
					\midrule
					\multicolumn{14}{c}{\textrm{Maxwell data}} \\ \midrule
					\text{ML} & 1 & 585 & 230 & 567 & 306 & 375 & 159 & 366 & 143 & 198 & 114 & 6442.96 & 6536.88 \\
					& 2 & 594 & 250 & 648 & 350 & 442 & 227 & 649 & 359 & 323 & 381 & 5954.08 & 6090.25 \\
					& 3 & 631 & 384 & 694 & 356 & 586 & 228 & 664 & 359 & 325 & 378 & 5893.17 & 6066.92 \\
					& 4 & 603 & 370 & 702 & 356 & 676 & 199 & 725 & \textbf{27024} & 276 & 391 & 5842.66 & 6049.27 \\ \midrule
					\text{Akaike[$n^{-1/2}$]} & 1 & 584 & 230 & 566 & 306 & 374 & 159 & 365 & 143 & 198 & 114 & 6442.96 & 6536.88 \\
					& 2 & 593 & 250 & 647 & 350 & 442 & 226 & 648 & 359 & 322 & 380 & 5954.08 & 6090.25 \\
					& 3 & 631 & 384 & 692 & 356 & 584 & 228 & 663 & 359 & 325 & 378 & 5893.17 & 6066.92 \\
					& 4 & 619 & 382 & 695 & 363 & 637 & 225 & 707 & 903 & 314 & 405 & 5847.36 & 6053.98 \\ \midrule
					\text{Hirose[$n^{-1/2}$]} & 1 & 584 & 230 & 567 & 306 & 374 & 159 & 366 & 143 & 198 & 114 & 6442.96 & 6536.88 \\
					& 2 & 594 & 250 & 648 & 350 & 442 & 227 & 649 & 359 & 323 & 381 & 5954.08 & 6090.25 \\
					& 3 & 631 & 384 & 693 & 356 & 585 & 228 & 663 & 359 & 325 & 378 & 5893.17 & 6066.92 \\
					& 4 & 620 & 383 & 696 & 364 & 638 & 225 & 708 & 905 & 315 & 406 & 5847.36 & 6053.98 \\
					\bottomrule
				\end{tabular}
\end{small}
\end{table}

\section{Concluding remarks} \label{sec:fapl_sum}

In this paper, we introduced a novel maximum softly penalised
likelihood framework for factor analysis models to address improper
solutions known as Heywood cases that frequently occur in statistical
practice. Heywood cases can lead to unstable and inconclusive results
related to factor loading estimates and factor scores, as well as
inaccurate inferences and model selection. Our approach provides a
comprehensive blueprint for constructing penalties and scaling factors
that ensure the existence of estimators within the admissible
parameter space and avoid the proposed ad hoc solutions in the
literature. Our work focuses on exploratory factor analysis, but the
proposed estimator can also be applied in a confirmatory factor
analysis setting by enforcing additional constraints that
practitioners might wish to impose on the model (e.g. zero loadings,
equal error variances, etc.). 

We provide sufficient conditions for the existence of the MPL
estimator in factor analysis, together with the asymptotic properties
of consistency and asymptotic normality of the MPL estimators.
Additionally, we derive decay rates for the scaling of the penalty
function to ensure consistency and asymptotic normality of the MSPL
estimators, thus preserving the favourable asymptotic properties
expected by the maximum likelihood estimator. Through extensive
simulation studies, we compared MSPL with appropriately scaled
versions of the penalties proposed by \citet{hirose+etal:2011}, which
are derived from Bayesian considerations and thus lack soft
penalisation by default. The MSPL estimators are found to recover the
performance expected from maximum likelihood theory while resolving
the issues related to Heywood cases, across various model
specifications, sample sizes, and item-to-factor ratios, making them a
valuable tool for practical applications in exploratory and
potentially confirmatory factor analysis. Our findings further reveal
that naive penalties not only can undermine frequentist properties, in
terms of higher bias, RMSE and probability of underestimation, but can
also have a deteriorating effect on the performance of model selection
procedures. Our framework enables
hypothesis testing by ensuring the existence of interior estimates in
finite samples, even when the ML estimator may fail to exist, while
preserving standard ML asymptotic behaviour.

A limitation of our results is that they assume the probability limit
of the sample covariance lies within the exploratory factor analysis
model class; extending the theory to a least-false interpretation
under covariance misspecification, in the spirit of
\citet{white:1982}, is a natural and important direction for future
work.  Further research directions include exploring alternative
penalty functions, within the MSPL framework, for the factor analysis
model~(\ref{eq:fa_model}) and for related models such as logistic
factor analysis \citep[see,][Chapter~4]{bartholomew.ea:11}.  For
example, in the logistic model, steep item characteristic curves lead
to infinite estimates of the loadings \citep[see, for example][for a
discussion of Heywood cases in item response models for binary
data]{wang.etal:23}. Maximum softly penalised likelihood can be
readily extended to handle those cases, too. 

\section{Supplementary material}

The supplementary material is available at
\url{https://github.com/psterzinger/FAPL}, and consists of the three
folders ``code'', ``results'', ``figures'', and the supplementary material
document. The latter provides the proofs to our results and evidence
from additional simulation studies and numerical examples to those
presented in the main text. The ``code'' directory contains scripts to
reproduce the numerical analyses, simulations, graphics and tables in
the main text and the supplementary material document. The ``results''
and ``figures'' directories provide all results and figures from the
numerical experiments and analyses in the main text and the
Supplementary Material, respectively.

\section*{Declarations}

For the purpose of open access, the authors have applied a Creative
Commons Attribution (CC BY) license to any Author Accepted Manuscript
version arising from this submission.

\bibliographystyle{chicago}
\bibliography{fapl-bib}



\section*{Supplementary material}

## S1.2 Data examples

Table S1: Estimates of $\mathbf{\Lambda}$ and $\mathbf{\Psi}$ for the Davis data using ML and MSPL with Akaike$[n^{-1/2}]$ and Hirose$[n^{-1/2}]$ penalties. A value 0.00 indicates a positive estimate that is less than 0.01, and $-0.00$ indicates a negative estimate that is greater than $-0.01$.

| $q$ | Item | ML | | | Akaike$[n^{-1/2}]$ | | | Hirose$[n^{-1/2}]$ | | |
|---|---|---|---|---|---|---|---|---|---|---|
| | | $\mathbf{\Lambda}_{\bullet,1}$ | $\mathbf{\Lambda}_{\bullet,2}$ | $\mathbf{\Psi}$ | $\mathbf{\Lambda}_{\bullet,1}$ | $\mathbf{\Lambda}_{\bullet,2}$ | $\mathbf{\Psi}$ | $\mathbf{\Lambda}_{\bullet,1}$ | $\mathbf{\Lambda}_{\bullet,2}$ | $\mathbf{\Psi}$ |
| 1 | 1 | $-0.81$ | | 0.34 | $-0.81$ | | 0.34 | $-0.81$ | | 0.34 |
| | 2 | $-0.81$ | | 0.34 | $-0.81$ | | 0.34 | $-0.81$ | | 0.34 |
| | 3 | $-0.48$ | | 0.77 | $-0.48$ | | 0.77 | $-0.48$ | | 0.77 |
| | 4 | $-0.41$ | | 0.83 | $-0.41$ | | 0.83 | $-0.41$ | | 0.83 |
| | 5 | $-0.67$ | | 0.55 | $-0.67$ | | 0.55 | $-0.67$ | | 0.55 |
| | 6 | $-0.89$ | | 0.20 | $-0.89$ | | 0.20 | $-0.89$ | | 0.20 |
| | 7 | $-0.84$ | | 0.29 | $-0.84$ | | 0.29 | $-0.84$ | | 0.30 |
| | 8 | $-0.66$ | | 0.57 | $-0.66$ | | 0.57 | $-0.66$ | | 0.57 |
| | 9 | $-0.84$ | | 0.30 | $-0.84$ | | 0.30 | $-0.84$ | | 0.30 |
| 2 | 1 | $-0.00$ | 3.76 | $-13.14$ | $-0.83$ | 0.43 | 0.12 | $-0.83$ | 0.43 | 0.12 |
| | 2 | $-0.77$ | 0.19 | 0.37 | $-0.81$ | 0.10 | 0.34 | $-0.81$ | 0.10 | 0.34 |
| | 3 | $-0.46$ | 0.11 | 0.78 | $-0.48$ | 0.02 | 0.77 | $-0.48$ | 0.02 | 0.77 |
| | 4 | $-0.41$ | 0.07 | 0.82 | $-0.41$ | $-0.14$ | 0.81 | $-0.41$ | $-0.13$ | 0.81 |
| | 5 | $-0.67$ | 0.14 | 0.54 | $-0.67$ | $-0.11$ | 0.53 | $-0.68$ | $-0.11$ | 0.53 |
| | 6 | $-0.88$ | 0.19 | 0.19 | $-0.89$ | $-0.10$ | 0.19 | $-0.90$ | $-0.09$ | 0.19 |
| | 7 | $-0.82$ | 0.18 | 0.30 | $-0.83$ | $-0.05$ | 0.30 | $-0.84$ | $-0.04$ | 0.30 |
| | 8 | $-0.65$ | 0.14 | 0.56 | $-0.66$ | $-0.09$ | 0.56 | $-0.66$ | $-0.09$ | 0.56 |
| | 9 | $-0.82$ | 0.18 | 0.30 | $-0.83$ | $-0.05$ | 0.30 | $-0.84$ | $-0.04$ | 0.30 |

Table S2: Estimates of $\mathbf{\Lambda}$ and $\mathbf{\Psi}$ for the Emmett data using ML and MSPL with Akaike$[n^{-1/2}]$ and Hirose$[n^{-1/2}]$ penalties. A value 0.00 indicates a positive estimate that is less than 0.01, and $-0.00$ indicates a negative estimate that is greater than $-0.01$.

| q | Item | ML | | | | | | Akaike$[n^{-1/2}]$ | | | | | | Hirose$[n^{-1/2}]$ | | | | | |
|---|---|---|---|---|---|---|---|---|---|---|---|---|---|---|---|---|---|---|---|
| | | $\Lambda_{\bullet,1}$ | $\Lambda_{\bullet,2}$ | $\Lambda_{\bullet,3}$ | $\Lambda_{\bullet,4}$ | $\Lambda_{\bullet,5}$ | $\Psi$ | $\Lambda_{\bullet,1}$ | $\Lambda_{\bullet,2}$ | $\Lambda_{\bullet,3}$ | $\Lambda_{\bullet,4}$ | $\Lambda_{\bullet,5}$ | $\Psi$ | $\Lambda_{\bullet,1}$ | $\Lambda_{\bullet,2}$ | $\Lambda_{\bullet,3}$ | $\Lambda_{\bullet,4}$ | $\Lambda_{\bullet,5}$ | $\Psi$ |
| 1 | 1 | −0.71 | | | | | 0.49 | −0.71 | | | | | 0.49 | −0.71 | | | | | 0.49 |
| | 2 | −0.73 | | | | | 0.46 | −0.73 | | | | | 0.46 | −0.73 | | | | | 0.46 |
| | 3 | −0.55 | | | | | 0.70 | −0.55 | | | | | 0.70 | −0.55 | | | | | 0.70 |
| | 4 | −0.74 | | | | | 0.45 | −0.74 | | | | | 0.45 | −0.74 | | | | | 0.45 |
| | 5 | −0.62 | | | | | 0.61 | −0.62 | | | | | 0.61 | −0.62 | | | | | 0.61 |
| | 6 | −0.69 | | | | | 0.52 | −0.69 | | | | | 0.52 | −0.69 | | | | | 0.52 |
| | 7 | −0.72 | | | | | 0.48 | −0.72 | | | | | 0.48 | −0.72 | | | | | 0.48 |
| | 8 | −0.47 | | | | | 0.78 | −0.47 | | | | | 0.78 | −0.47 | | | | | 0.78 |
| | 9 | −0.82 | | | | | 0.33 | −0.81 | | | | | 0.34 | −0.82 | | | | | 0.34 |
| 2 | 1 | −0.73 | −0.11 | | | | 0.46 | −0.72 | −0.10 | | | | 0.46 | −0.73 | −0.10 | | | | 0.46 |
| | 2 | −0.73 | −0.04 | | | | 0.46 | −0.73 | −0.03 | | | | 0.46 | −0.73 | −0.03 | | | | 0.47 |
| | 3 | −0.56 | −0.14 | | | | 0.67 | −0.56 | −0.13 | | | | 0.67 | −0.56 | −0.13 | | | | 0.67 |
| | 4 | −0.72 | 0.54 | | | | 0.19 | −0.71 | 0.54 | | | | 0.19 | −0.71 | 0.55 | | | | 0.19 |
| | 5 | −0.58 | 0.50 | | | | 0.41 | −0.58 | 0.50 | | | | 0.41 | −0.58 | 0.51 | | | | 0.41 |
| | 6 | −0.67 | 0.58 | | | | 0.22 | −0.66 | 0.58 | | | | 0.22 | −0.66 | 0.59 | | | | 0.22 |
| | 7 | −0.75 | −0.18 | | | | 0.40 | −0.75 | −0.18 | | | | 0.40 | −0.75 | −0.18 | | | | 0.40 |
| | 8 | −0.49 | −0.11 | | | | 0.74 | −0.49 | −0.11 | | | | 0.74 | −0.49 | −0.11 | | | | 0.75 |
| | 9 | −0.86 | −0.19 | | | | 0.22 | −0.86 | −0.19 | | | | 0.22 | −0.86 | −0.18 | | | | 0.22 |
| 3 | 1 | −0.70 | −0.24 | 0.01 | | | 0.45 | −0.70 | −0.23 | 0.01 | | | 0.45 | −0.71 | −0.21 | 0.04 | | | 0.45 |
| | 2 | −0.73 | −0.06 | −0.20 | | | 0.43 | −0.72 | −0.06 | −0.21 | | | 0.43 | −0.73 | −0.06 | −0.19 | | | 0.43 |
| | 3 | −0.54 | −0.12 | −0.27 | | | 0.62 | −0.54 | −0.13 | −0.27 | | | 0.62 | −0.55 | −0.13 | −0.25 | | | 0.62 |
| | 4 | −0.78 | 0.38 | 0.18 | | | 0.21 | −0.77 | 0.40 | 0.15 | | | 0.21 | −0.77 | 0.43 | 0.13 | | | 0.21 |
| | 5 | −0.65 | 0.43 | 0.06 | | | 0.38 | −0.64 | 0.44 | 0.04 | | | 0.38 | −0.64 | 0.46 | 0.00 | | | 0.38 |
| | 6 | −0.74 | 0.40 | 0.33 | | | 0.18 | −0.74 | 0.42 | 0.31 | | | 0.18 | −0.73 | 0.46 | 0.28 | | | 0.18 |
| | 7 | −0.72 | −0.24 | −0.16 | | | 0.40 | −0.71 | −0.24 | −0.16 | | | 0.40 | −0.73 | −0.23 | −0.13 | | | 0.40 |
| | 8 | −0.47 | −0.41 | 0.38 | | | 0.46 | −0.48 | −0.39 | 0.39 | | | 0.46 | −0.49 | −0.34 | 0.42 | | | 0.47 |
| | 9 | −0.82 | −0.29 | −0.10 | | | 0.23 | −0.82 | −0.28 | −0.10 | | | 0.23 | −0.83 | −0.26 | −0.06 | | | 0.23 |
| 4 | 1 | −0.56 | −0.16 | −0.10 | −0.44 | | 0.46 | −0.69 | −0.23 | 0.05 | −0.10 | | 0.45 | −0.70 | −0.20 | 0.09 | −0.05 | | 0.45 |
| | 2 | −0.47 | −0.00 | −0.13 | −0.57 | | 0.44 | −0.70 | −0.11 | −0.18 | −0.14 | | 0.44 | −0.72 | −0.12 | −0.16 | −0.04 | | 0.44 |
| | 3 | −0.01 | −0.00 | −3.78 | 0.02 | | −13.31 | −0.56 | −0.23 | −0.45 | 0.18 | | 0.40 | −0.56 | −0.25 | −0.34 | 0.33 | | 0.40 |
| | 4 | −0.70 | 0.46 | −0.09 | −0.30 | | 0.20 | −0.79 | 0.40 | 0.02 | 0.05 | | 0.20 | −0.78 | 0.43 | 0.00 | 0.07 | | 0.20 |
| | 5 | −0.53 | 0.48 | −0.07 | −0.31 | | 0.39 | −0.65 | 0.43 | −0.04 | −0.08 | | 0.38 | −0.65 | 0.44 | −0.09 | −0.05 | | 0.38 |
| | 6 | −0.73 | 0.47 | −0.07 | −0.19 | | 0.20 | −0.75 | 0.45 | 0.17 | 0.05 | | 0.19 | −0.74 | 0.49 | 0.14 | 0.03 | | 0.19 |
| | 7 | −0.49 | −0.18 | −0.12 | −0.56 | | 0.40 | −0.69 | −0.27 | −0.07 | −0.19 | | 0.40 | −0.72 | −0.27 | −0.05 | −0.11 | | 0.40 |
| | 8 | −0.73 | −0.43 | −0.05 | 0.12 | | 0.26 | −0.51 | −0.35 | 0.42 | 0.23 | | 0.38 | −0.49 | −0.26 | 0.53 | 0.17 | | 0.38 |
| | 9 | −0.60 | −0.21 | −0.13 | −0.59 | | 0.23 | −0.80 | −0.30 | −0.03 | −0.17 | | 0.23 | −0.83 | −0.29 | 0.00 | −0.09 | | 0.23 |
| 5 | 1 | −0.67 | −0.26 | 0.08 | 0.15 | 0.05 | 0.45 | −0.70 | −0.22 | 0.08 | 0.04 | −0.04 | 0.45 | −0.70 | −0.22 | 0.08 | 0.05 | −0.03 | 0.45 |
| | 2 | −0.66 | −0.20 | −0.16 | 0.16 | 0.34 | 0.35 | −0.73 | −0.12 | −0.19 | −0.07 | 0.22 | 0.36 | −0.73 | −0.13 | −0.20 | −0.06 | 0.22 | 0.36 |
| | 3 | −0.51 | −0.20 | −0.20 | 0.11 | 0.15 | 0.62 | −0.54 | −0.21 | −0.25 | 0.07 | 0.06 | 0.59 | −0.54 | −0.22 | −0.25 | 0.07 | 0.06 | 0.59 |
| | 4 | −0.03 | 0.00 | −0.00 | 2.98 | 0.00 | −7.88 | −0.78 | 0.42 | −0.01 | 0.19 | −0.03 | 0.17 | −0.79 | 0.42 | −0.01 | 0.20 | −0.02 | 0.17 |
| | 5 | −0.57 | 0.33 | 0.08 | 0.23 | 0.25 | 0.45 | −0.64 | 0.46 | −0.10 | −0.10 | −0.11 | 0.34 | −0.65 | 0.45 | −0.11 | −0.08 | −0.11 | 0.34 |
| | 6 | −0.63 | 0.35 | 0.46 | 0.26 | 0.44 | 0.00 | −0.74 | 0.49 | 0.19 | 0.02 | 0.08 | 0.16 | −0.74 | 0.49 | 0.18 | 0.03 | 0.09 | 0.17 |
| | 7 | −0.78 | −0.17 | −0.10 | 0.14 | −0.12 | 0.32 | −0.72 | −0.28 | −0.08 | −0.10 | −0.19 | 0.35 | −0.73 | −0.28 | −0.07 | −0.08 | −0.19 | 0.35 |
| | 8 | −0.43 | −0.39 | 0.42 | 0.09 | −0.08 | 0.47 | −0.48 | −0.26 | 0.43 | 0.10 | −0.04 | 0.50 | −0.48 | −0.25 | 0.43 | 0.11 | −0.03 | 0.50 |
| | 9 | −0.79 | −0.31 | −0.00 | 0.16 | 0.10 | 0.24 | −0.82 | −0.29 | 0.01 | −0.09 | 0.02 | 0.23 | −0.82 | −0.29 | 0.02 | −0.08 | 0.03 | 0.23 |

Table S3: Estimates of $\mathbf{\Lambda}$ and $\mathbf{\Psi}$ for the Maxwell data using ML and MSPL with Akaike$[n^{-1/2}]$ and Hirose$[n^{-1/2}]$ penalties. A value 0.00 indicates a positive estimate that is less than 0.01, and $-0.00$ indicates a negative estimate that is greater than $-0.01$.

| $q$ | Item | ML | | | | | Akaike$[n^{-1/2}]$ | | | | | Hirose$[n^{-1/2}]$ | | | | |
|---|---|---|---|---|---|---|---|---|---|---|---|---|---|---|---|---|
| | | $\mathbf{\Lambda}_{\bullet,1}$ | $\mathbf{\Lambda}_{\bullet,2}$ | $\mathbf{\Lambda}_{\bullet,3}$ | $\mathbf{\Lambda}_{\bullet,4}$ | $\mathbf{\Psi}$ | $\mathbf{\Lambda}_{\bullet,1}$ | $\mathbf{\Lambda}_{\bullet,2}$ | $\mathbf{\Lambda}_{\bullet,3}$ | $\mathbf{\Lambda}_{\bullet,4}$ | $\mathbf{\Psi}$ | $\mathbf{\Lambda}_{\bullet,1}$ | $\mathbf{\Lambda}_{\bullet,2}$ | $\mathbf{\Lambda}_{\bullet,3}$ | $\mathbf{\Lambda}_{\bullet,4}$ | $\mathbf{\Psi}$ |
| 1 | 1 | −0.76 | | | | 0.42 | −0.76 | | | | 0.42 | −0.76 | | | | 0.42 |
| | 2 | −0.48 | | | | 0.77 | −0.48 | | | | 0.77 | −0.48 | | | | 0.77 |
| | 3 | −0.75 | | | | 0.43 | −0.75 | | | | 0.43 | −0.75 | | | | 0.43 |
| | 4 | −0.55 | | | | 0.69 | −0.55 | | | | 0.69 | −0.55 | | | | 0.69 |
| | 5 | −0.61 | | | | 0.63 | −0.61 | | | | 0.63 | −0.61 | | | | 0.63 |
| | 6 | −0.40 | | | | 0.84 | −0.40 | | | | 0.84 | −0.40 | | | | 0.84 |
| | 7 | −0.60 | | | | 0.63 | −0.60 | | | | 0.63 | −0.60 | | | | 0.63 |
| | 8 | −0.38 | | | | 0.86 | −0.38 | | | | 0.86 | −0.38 | | | | 0.86 |
| | 9 | −0.45 | | | | 0.80 | −0.45 | | | | 0.80 | −0.45 | | | | 0.80 |
| | 10 | −0.34 | | | | 0.89 | −0.34 | | | | 0.89 | −0.34 | | | | 0.89 |
| 2 | 1 | −0.75 | 0.18 | | | 0.41 | −0.75 | 0.18 | | | 0.41 | −0.77 | 0.06 | | | 0.41 |
| | 2 | −0.47 | 0.18 | | | 0.75 | −0.47 | 0.18 | | | 0.75 | −0.49 | 0.11 | | | 0.75 |
| | 3 | −0.75 | 0.29 | | | 0.35 | −0.75 | 0.29 | | | 0.35 | −0.79 | 0.17 | | | 0.35 |
| | 4 | −0.54 | 0.25 | | | 0.65 | −0.54 | 0.25 | | | 0.65 | −0.57 | 0.16 | | | 0.65 |
| | 5 | −0.60 | 0.30 | | | 0.56 | −0.59 | 0.30 | | | 0.56 | −0.63 | 0.20 | | | 0.56 |
| | 6 | −0.41 | −0.23 | | | 0.77 | −0.41 | −0.23 | | | 0.77 | −0.37 | −0.30 | | | 0.77 |
| | 7 | −0.68 | −0.43 | | | 0.35 | −0.68 | −0.43 | | | 0.35 | −0.60 | −0.54 | | | 0.35 |
| | 8 | −0.42 | −0.43 | | | 0.64 | −0.42 | −0.43 | | | 0.64 | −0.35 | −0.49 | | | 0.64 |
| | 9 | −0.47 | −0.31 | | | 0.68 | −0.47 | −0.31 | | | 0.68 | −0.42 | −0.38 | | | 0.68 |
| | 10 | −0.39 | −0.48 | | | 0.62 | −0.39 | −0.48 | | | 0.62 | −0.31 | −0.53 | | | 0.62 |
| 3 | 1 | −0.73 | 0.26 | 0.20 | | 0.37 | −0.72 | 0.25 | 0.22 | | 0.37 | −0.72 | 0.24 | 0.22 | | 0.37 |
| | 2 | −0.47 | 0.23 | −0.34 | | 0.62 | −0.48 | 0.23 | −0.32 | | 0.62 | −0.48 | 0.23 | −0.32 | | 0.62 |
| | 3 | −0.73 | 0.37 | −0.15 | | 0.31 | −0.74 | 0.37 | −0.12 | | 0.31 | −0.74 | 0.36 | −0.12 | | 0.31 |
| | 4 | −0.51 | 0.29 | −0.10 | | 0.64 | −0.52 | 0.29 | −0.08 | | 0.64 | −0.52 | 0.29 | −0.08 | | 0.64 |
| | 5 | −0.57 | 0.39 | 0.34 | | 0.41 | −0.56 | 0.37 | 0.36 | | 0.42 | −0.56 | 0.37 | 0.36 | | 0.42 |
| | 6 | −0.43 | −0.20 | −0.03 | | 0.77 | −0.43 | −0.20 | −0.02 | | 0.77 | −0.43 | −0.20 | −0.02 | | 0.77 |
| | 7 | −0.72 | −0.38 | 0.10 | | 0.34 | −0.71 | −0.38 | 0.11 | | 0.34 | −0.71 | −0.39 | 0.11 | | 0.34 |
| | 8 | −0.46 | −0.39 | −0.04 | | 0.64 | −0.46 | −0.39 | −0.04 | | 0.64 | −0.46 | −0.39 | −0.04 | | 0.64 |
| | 9 | −0.50 | −0.27 | −0.04 | | 0.67 | −0.50 | −0.27 | −0.03 | | 0.67 | −0.50 | −0.27 | −0.03 | | 0.68 |
| | 10 | −0.43 | −0.44 | 0.00 | | 0.62 | −0.43 | −0.44 | 0.00 | | 0.62 | −0.43 | −0.44 | 0.00 | | 0.62 |
| 4 | 1 | −0.74 | 0.14 | 0.19 | 0.03 | 0.40 | −0.72 | 0.20 | 0.24 | −0.11 | 0.38 | −0.72 | 0.19 | 0.22 | −0.12 | 0.38 |
| | 2 | −0.46 | 0.32 | −0.24 | 0.02 | 0.63 | −0.50 | 0.23 | −0.28 | 0.00 | 0.62 | −0.50 | 0.22 | −0.29 | 0.00 | 0.62 |
| | 3 | −0.74 | 0.39 | −0.06 | 0.04 | 0.30 | −0.76 | 0.33 | −0.07 | −0.00 | 0.30 | −0.77 | 0.32 | −0.08 | −0.01 | 0.30 |
| | 4 | −0.51 | 0.30 | 0.00 | 0.03 | 0.64 | −0.54 | 0.26 | −0.01 | 0.09 | 0.64 | −0.54 | 0.25 | −0.02 | 0.09 | 0.64 |
| | 5 | −0.61 | 0.21 | 0.51 | 0.03 | 0.32 | −0.57 | 0.31 | 0.46 | 0.02 | 0.36 | −0.58 | 0.31 | 0.45 | 0.02 | 0.36 |
| | 6 | −0.40 | −0.14 | −0.11 | 0.07 | 0.80 | −0.42 | −0.21 | −0.01 | 0.04 | 0.77 | −0.42 | −0.22 | −0.02 | 0.03 | 0.77 |
| | 7 | −0.72 | −0.43 | −0.12 | 0.09 | 0.28 | −0.68 | −0.42 | 0.08 | −0.25 | 0.29 | −0.67 | −0.43 | 0.08 | −0.26 | 0.29 |
| | 8 | −0.00 | −0.00 | −0.00 | 5.20 | −26.02 | −0.50 | −0.58 | 0.02 | 0.56 | 0.10 | −0.50 | −0.59 | 0.03 | 0.55 | 0.10 |
| | 9 | −0.46 | −0.20 | −0.14 | 0.08 | 0.72 | −0.48 | −0.29 | −0.02 | 0.03 | 0.69 | −0.48 | −0.29 | −0.02 | 0.03 | 0.69 |
| | 10 | −0.42 | −0.42 | −0.20 | 0.06 | 0.61 | −0.39 | −0.44 | −0.04 | −0.24 | 0.59 | −0.38 | −0.44 | −0.04 | −0.25 | 0.59 |

## S2 Proofs

### S2.1 Existence

We start by stating an existence master theorem, which we use to establish our existence results for MPL in factor analysis.

**Theorem A1** (Existence master theorem). *Let $\mathcal{X} \subseteq \Re^d$, denote by $\mathrm{cl}(\mathcal{X})$ its closure and let $\partial \mathcal{X} \subseteq \mathrm{cl}(\mathcal{X})$ be a set of sequential limit points of $\mathcal{X}$ and denote $\mathrm{int}(\mathcal{X}) = \mathrm{cl}(\mathcal{X}) \setminus \partial \mathcal{X}$.*

*Let $h : \mathcal{X} \to \Re$ be a function such that*

*A1) $h(x)$ is continuous on $\mathcal{X}$*

*A2) $\sup\limits_{x \in \mathcal{X}} h(x) < \infty$*

A3) For any sequence $\{x_n\}_{n \in \mathbb{N}}$, $x_n \in \mathcal{X}$ such that either $\lim_{n \to \infty} x_n \in \partial\mathcal{X}$,[1] or $\|x_n\| \to \infty$, $\lim_{n \to \infty} h(x_n) = -\infty$

Then the set of maximisers is nonempty, i.e.

$$\left\{ x^* \in \text{int}(\mathcal{X}) : h(x^*) = \sup_{x \in \mathcal{X}} h(x) \right\} \neq \emptyset.$$

*Proof.* Towards a contradiction, assume that supremum of $h$ is not attained in $\text{int}(\mathcal{X})$, that is, for all $x \in \text{int}(\mathcal{X})$, $h(x) < H^*$, where $H^* = \sup_{x \in \mathcal{X}} h(x)$. Note that by assumption A2), $H^* \in \Re$.

Construct a sequence $\{x_n\}_{n \in \mathbb{N}}$, $x_n \in \mathcal{X}$ such that $\lim_{n \to \infty} h(x_n) = H^*$. To do this, note that for any $\varepsilon > 0$, one can find a $x \in \mathcal{X}$ for which $H^* - h(x) < \varepsilon$. If this were not the case then, there must exist a $\varepsilon > 0$ such that for all $x \in \mathcal{X}$, $H^* - \varepsilon > h(x)$, contradicting the least upper bound property of $H^*$. Thus, one can construct $\{x_n\}_{n \in \mathbb{N}}$ by choosing any element of the set $\{x \in \mathcal{X} : H^* - h(x) < 1/n\}$.

Next, note that $\{x_n\}_{n \in \mathbb{N}}$ must be bounded. For this, assume that on the contrary $\|x_n\| \to \infty$ as $n \to \infty$. Then by A3), $h(x_n) \to -\infty$ as $n \to \infty$. But then for every $r \in \Re$ there exists a $N \in \mathbb{N}$ such that for all $n > N$, $h(x_n) < r$. But this stands in contradiction to the construction of $\{x_n\}_{n \in \mathbb{N}}$, for which $\{h(x_n)\}_{n \in \mathbb{N}}$ ought to converge to $H^* \in \Re$.

Then, by the Bolzano-Weierstrass theorem (see for example Rudin 1976, Theorem 3.6 (b)), $\{x_n\}_{n \in \mathbb{N}}$ must contain a convergent subsequence, say $\{x_{n_k}\}_{k \in \mathbb{N}}$, with limit $x^* \in \text{cl}(\mathcal{X})$ and where $\{n_k\}_{k \in \mathbb{N}} \in \mathbb{N} : n_s < n_t$ for $s < t$. Now, by construction of $\text{int}(\mathcal{X}) = \text{cl}(\mathcal{X}) \setminus \partial\mathcal{X}$, one of the the two cases below must hold.

(i) $x^* \in \text{int}(\mathcal{X})$: Then, using that the subsequential limit of a convergent sequence must equal the limit of that sequence (e.g. Rudin 1976, Definition 3.5), it follows that

$$H^* = \lim_{k \to \infty} h(x_{n_k}) = h\left(\lim_{k \to \infty} x_{n_k}\right) = h(x^*),$$

where the second equality follows from continuity of $h(x)$, which holds by A1). But then there exists a $x^* \in \text{int}(\mathcal{X})$ such that $h(x^*) = H^* = \sup_{x \in \mathcal{X}} h(x)$ which was assumed not to be case.

(ii) $x^* \in \partial\mathcal{X}$: Then

$$H^* = h(x^*) = h\left(\lim_{k \to \infty} x_{n_k}\right) = \lim_{k \to \infty} h(x_{n_k}) = -\infty,$$

where the third equality follows since $h(x)$ is continuous on $\mathcal{X}$ (see for example Rudin 1976, Theorems 4.6-4.7) which holds by A1) and the last from the decay condition A3). This stands in contradiction to $H^* \in \Re$.

Since either case leads to a contradiction, the initial assumption must be false. Thus, there must exists a $x^* \in \text{int}(\mathcal{X}) : h(x^*) = H^*$.

□

Next, we state a general existence theorem for factor analysis. This result is more general than the existence Theorem 1 stated in the main text. In particular, it allows optimisation to be conducted in a general parameter space $\boldsymbol{\Theta} \subseteq \Re^d$, to allow for any potential constraints that one wishes to impose on the optimisation problem. Each vector $\boldsymbol{\theta} \in \boldsymbol{\Theta}$ is mapped to a symmetric, positive definite, $p \times p$ matrix through the map $\boldsymbol{\theta} \mapsto s(\boldsymbol{\theta})$.

**Theorem A2.** *Let $\boldsymbol{\Theta} \subseteq \Re^d$, denote by $\text{cl}(\boldsymbol{\Theta})$ its closure, $\partial\boldsymbol{\Theta} \subseteq \text{cl}(\boldsymbol{\Theta})$ a set of sequential limit points of $\boldsymbol{\Theta}$, and denote $\text{int}(\boldsymbol{\Theta}) = \text{cl}(\boldsymbol{\Theta}) \setminus \partial(\boldsymbol{\Theta})$.*

*Let $s : \boldsymbol{\theta} \mapsto s(\boldsymbol{\theta})$ be a mapping from $\boldsymbol{\Theta}$ to the space of $p \times p$ positive definite and symmetric matrices and define $\ell^*(\boldsymbol{\theta}; \boldsymbol{S}) = \ell(s(\boldsymbol{\theta}); \boldsymbol{S}) + P^*(\boldsymbol{\theta})$, where $\ell(\boldsymbol{\Sigma}; \boldsymbol{S})$ is the profile log-likelihood of the EFA model in Section 2 and $\boldsymbol{S}$ is full rank.*

*Assume that:*

---

[1] Formally: $\lim_{n \to \infty} \inf_{x \in \partial\mathcal{X}} \|x_n - x\| = 0$.

6

*A4)* $s$ and $P^*$ are continuous on $\Theta$

*A5)* For any sequence $\{\boldsymbol{\theta}_n\}_{n\in\mathbb{N}}$, $\boldsymbol{\theta}_n \in \Theta$, such that $\|\boldsymbol{\theta}_n\| \to \infty$ as $n \to \infty$, either $P^*(\boldsymbol{\theta}) \to -\infty$ or $\|\|s(\boldsymbol{\theta}_n)\|\| \to \infty$

*A6)* For any sequence $\{\boldsymbol{\theta}_n\}_{n\in\mathbb{N}}$, $\boldsymbol{\theta}_n \in \Theta$, such that (i) $\boldsymbol{\theta}_n \in \partial\Theta$ as $n \to \infty$, and (ii) $\lambda_{\min}(s(\boldsymbol{\theta}_n)) \not\to 0$, $P^*(\boldsymbol{\theta}_n) \to -\infty$

*A7)* $\sup_{\boldsymbol{\theta}\in\Theta} P^*(\boldsymbol{\theta}) < \infty$

*then the set of maximisers is nonempty, i.e.*

$$\left\{\boldsymbol{\theta}^* \in \mathrm{int}(\Theta) : \ell^*(\boldsymbol{\theta}^*; \boldsymbol{S}) = \sup_{\boldsymbol{\theta}\in\Theta} \ell^*(\boldsymbol{\theta}; \boldsymbol{S})\right\} \neq \emptyset.$$

*Proof.* We verify that the conditions of Theorem A1 are met for $\ell^*(\boldsymbol{\theta}; \boldsymbol{S})$.

A1) It is evident that $\ell(\boldsymbol{\Sigma}; \boldsymbol{S})$ is continuous on the space of $p \times p$ positive definite, symmetric matrices. By assumption A4), the map $\boldsymbol{\theta} \mapsto s(\boldsymbol{\theta})$ is continuous so that $\ell(s(\boldsymbol{\theta}); \boldsymbol{S})$ is continuous. Additionally, by A4), $P^*(\boldsymbol{\theta})$ is continuous in $\boldsymbol{\theta}$, so that $\ell^*(\boldsymbol{\theta}; \boldsymbol{S})$ is continuous in $\boldsymbol{\theta}$ on $\Theta$.

A2) Next, let $\mathcal{S}$ be the space of $p \times p$ symmetric nonnegative definite matrices. Burg et al. (1982, Section IV) show that if $\boldsymbol{S}$ is full rank, then $\boldsymbol{S}$ is the unique maximiser of $\ell(\boldsymbol{\Sigma}; \boldsymbol{S})$ over $\mathcal{S}$. Now note that $\mathcal{S}_{\boldsymbol{\theta}} = \{\boldsymbol{\Sigma} : \boldsymbol{\Sigma} = s(\boldsymbol{\theta}), \boldsymbol{\theta} \in \Theta\} \subseteq \mathcal{S}$. Hence,

$$\begin{aligned}
\sup_{\boldsymbol{\theta}\in\Theta} \ell^*(\boldsymbol{\theta}; \boldsymbol{S}) &= \sup_{\boldsymbol{\theta}\in\Theta} \{\ell(s(\boldsymbol{\theta}); \boldsymbol{S}) + P^*(\boldsymbol{\theta})\} \\
&\leq \sup_{\boldsymbol{\Sigma}\in\mathcal{S}_{\boldsymbol{\theta}}} \ell(\boldsymbol{\Sigma}; \boldsymbol{S}) + \sup_{\boldsymbol{\theta}\in\Theta} P^*(\boldsymbol{\theta}) \\
&\leq \sup_{\boldsymbol{\Sigma}\in\mathcal{S}} \ell(\boldsymbol{\Sigma}; \boldsymbol{S}) + \sup_{\boldsymbol{\theta}\in\Theta} P^*(\boldsymbol{\theta}) \\
&= \ell(\boldsymbol{S}; \boldsymbol{S}) + \sup_{\boldsymbol{\theta}\in\Theta} P^*(\boldsymbol{\theta}) \\
&\leq \ell(\boldsymbol{S}; \boldsymbol{S}) + C_p \\
&< \infty,
\end{aligned}$$

where the boundedness of $P^*(\boldsymbol{\theta})$ from above comes from assumption A7).

A3) Consider a sequence $\{\boldsymbol{\theta}_n\}_{n\in\mathbb{N}}$. (i) Assume that $\boldsymbol{\theta}_n \in \Theta$ and $\lambda_{\min}(s(\boldsymbol{\theta}_n)) \not\to 0$, as $n \to \infty$. Then

$$\begin{aligned}
\ell^*(\boldsymbol{\theta}_n; \boldsymbol{S}) &= \ell(s(\boldsymbol{\theta}_n); \boldsymbol{S}) + P^*(\boldsymbol{\theta}_n) \\
&\leq \sup_{\boldsymbol{\Sigma}\in\mathcal{S}} \{\ell(\boldsymbol{\Sigma}; \boldsymbol{S})\} + P^*(\boldsymbol{\theta}_n) \\
&= \ell(\boldsymbol{S}; \boldsymbol{S}) + P^*(\boldsymbol{\theta}_n) \\
&\to -\infty \quad \text{as } n \to \infty,
\end{aligned} \tag{S1}$$

where the last line follows from A5). (ii) Assume that $\boldsymbol{\theta}_n \in \Theta$ and $\lambda_{\min}(s(\boldsymbol{\theta}_n)) \to 0$, as $n \to \infty$. Then Burg et al. (1982, Section II) show that $\ell(s(\boldsymbol{\theta}_n); \boldsymbol{S}) \to -\infty$ as $n \to \infty$. By A7), $\ell^*(s(\boldsymbol{\theta}_n); \boldsymbol{S}) \to -\infty$ as $n \to \infty$. (iii) Assume that $\|\boldsymbol{\theta}_n\| \to \infty$ as $n \to \infty$. If $P^*(\boldsymbol{\theta}_n) \to \infty$, then the bound from (S1) establishes that $\ell^*(\boldsymbol{\theta}_n; \boldsymbol{S}) \to -\infty$. If on the other hand $\|\|s(\boldsymbol{\theta}_n)\|\| \to \infty$, Burg et al. (1982, Section II) show that $\ell(s(\boldsymbol{\theta}_n); \boldsymbol{S}) \to -\infty$ as $n \to \infty$. Therefore

$$\begin{aligned}
\ell^*(\boldsymbol{\theta}_n; \boldsymbol{S}) &= \ell(s(\boldsymbol{\theta}_n); \boldsymbol{S}) + P^*(\boldsymbol{\theta}_n) \\
&\leq \ell(s(\boldsymbol{\theta}_n); \boldsymbol{S})\} + \sup_{\boldsymbol{\theta}\in\Theta}\{P^*(\boldsymbol{\theta}_n)\} \\
&\leq \ell(s(\boldsymbol{\theta}_n); \boldsymbol{S})\} + C_p \\
&\to -\infty \quad \text{as } n \to \infty,
\end{aligned}$$

From this result follows the existence of MPL estimates as stated in Theorem 1 in the main text.

**Theorem A3** (Existence of MPL estimates in factor analysis). *Let $\partial\Theta = \{\boldsymbol{\theta} \in \Re^{p(q+1)} : \exists m > pq, \theta_m = 0\}$ and $\boldsymbol{\Sigma}(\boldsymbol{\theta}) = \boldsymbol{\Lambda}(\boldsymbol{\theta})\boldsymbol{\Lambda}(\boldsymbol{\theta})^\top + \boldsymbol{\Psi}(\boldsymbol{\theta})$. Assume that the penalty function $P^*(\boldsymbol{\theta}) : \boldsymbol{\Theta} \to \Re$*

*E1) is continuous on $\boldsymbol{\Theta}$,*

*E2) is bounded from above on $\boldsymbol{\Theta}$, i.e. $\sup_{\boldsymbol{\theta}\in\boldsymbol{\Theta}} P^*(\boldsymbol{\theta}) < \infty$, and*

*E3) diverges to $-\infty$ for any sequence $\{\boldsymbol{\theta}(r)\}_{r\in\mathbb{N}}$ such that $\lim_{r\to\infty} \boldsymbol{\theta}(r) \in \partial\boldsymbol{\Theta}$ and $\lim_{r\to\infty} \lambda_{\min}(\boldsymbol{\Sigma}(\boldsymbol{\theta}(r))) > 0$.*

*Then, the set of maximum penalized likelihood estimates $\tilde{\boldsymbol{\theta}} \in \arg\max_{\boldsymbol{\theta}\in\boldsymbol{\Theta}} \{\ell^*(\boldsymbol{\theta}; S)\}$ is non-empty, when $\boldsymbol{S}$ has full rank.*

*Proof.* We verify that the conditions of Theorem A2 are met for $\boldsymbol{\Theta} = \{\boldsymbol{\theta} \in \Re^{p(q+1)} : \forall i > pq, \theta_i > 0\}$ and $\partial\boldsymbol{\Theta} = \{\boldsymbol{\theta} \in \Re^{p(q+1)} : \exists m > pq, \theta_m = 0\}$ under E1)-E3).

First note that $\partial\boldsymbol{\Theta}$ is the boundary of $\boldsymbol{\Theta}$ and $\text{int}(\boldsymbol{\Theta}) = \boldsymbol{\Theta}$. Further recall, that $\boldsymbol{\theta} \mapsto \boldsymbol{\Sigma}(\boldsymbol{\theta})$ for $\boldsymbol{\Sigma} = \boldsymbol{\Lambda}\boldsymbol{\Lambda}^\top + \boldsymbol{\Psi}$ for $\boldsymbol{\theta} = (\theta_1, \ldots, \theta_{p(q+1)})^\top = (\lambda_{11}, \ldots, \lambda_{pq}, \psi_{11}, \ldots, \psi_{pp})^\top$, where $\lambda_{jk}$ and $\psi_{jj}$ are the $(j,k)$th and $(j,j)$th elements of $\boldsymbol{\Lambda}$ and $\boldsymbol{\Psi}$, respectively ($j = 1, \ldots, p; k = 1, \ldots, q$). Thus, for each $\boldsymbol{\theta} \in \boldsymbol{\Theta}$, $\boldsymbol{\Sigma}$ is positive definite.

It remains to verify assumptions A4)-A7).

A4) It is readily seen that $\boldsymbol{\theta} \mapsto \boldsymbol{\Sigma}(\boldsymbol{\theta})$ is continuous on $\boldsymbol{\Theta}$. Further, $P^*$ is continuous by assumption E1).

A5) Let $\boldsymbol{\theta}(r)$ be a diverging sequence in $\boldsymbol{\Theta}$ as $r \to \infty$ and $\boldsymbol{\Sigma}(r)$ the associated sequence of variance-covariance matrices. The $i$th element of $\boldsymbol{\Sigma}(r)$ is given by
$$\boldsymbol{\Sigma}(r)_{ii} = \sum_{j=1} \lambda_{ij}(r)^2 + \psi_{ii}(r).$$

Hence if either $|\lambda_{ij}(r)| \to \infty$ or $\psi_{ii}(r) \to \infty$ as $r \to \infty$, also $\boldsymbol{\Sigma}_{ii}(r) \to \infty$ and consequently $\|\boldsymbol{\Sigma}(r)\| \to \infty$ as required.

A6) Holds by E3).

A7) Holds by E2).

This concludes the proof. □

## S2.2 Consistency

As we did for existence, and for the same reasons, we first provide a consistency result for general parameter spaces $\boldsymbol{\Theta}$. From this, the existence result of Theorem 2 follows as a corollary. These more general parameterisations are for example required for the $\sqrt{n}$-consistency results of Section 5.2 and they might be desirable if one wishes to impose further restrictions on the structure of $\boldsymbol{\Sigma}$.

**Theorem A4.** *Let $\boldsymbol{\Theta} \subseteq \Re^d$, denote by $cl(\boldsymbol{\Theta})$ its closure, $\partial\boldsymbol{\Theta}$ a set of sequential limit points of $\boldsymbol{\Theta}$ and denote by $int(\boldsymbol{\Theta}) = cl(\boldsymbol{\Theta}) \setminus \partial\boldsymbol{\Theta}$. Let $\boldsymbol{\Sigma}(\boldsymbol{\theta}) = \boldsymbol{\Lambda}(\boldsymbol{\theta})\boldsymbol{\Lambda}(\boldsymbol{\theta})^\top + \boldsymbol{\Psi}(\boldsymbol{\theta})$ for some maps $\boldsymbol{\theta} \mapsto \boldsymbol{\Lambda}(\boldsymbol{\theta})$, $\boldsymbol{\theta} \mapsto \boldsymbol{\Psi}(\boldsymbol{\theta})$. Finally, denote by $\ell^*(\boldsymbol{\theta}; \boldsymbol{S}) = \ell(\boldsymbol{\Sigma}(\boldsymbol{\theta}); \boldsymbol{S}) + P^*(\boldsymbol{\theta})$ the profile log-likelihood of the EFA model in Section 2.*

*Assume that*

*A8) the factor model is strongly identifiable*

*A9) There exists a $\boldsymbol{\theta}_0 \in int(\boldsymbol{\Theta})$ such that $\boldsymbol{\Sigma}(\boldsymbol{\theta}_0) = \boldsymbol{\Sigma}_0$*

8

*A10)*
$$\left\{\boldsymbol{\theta}^* \in \text{int}(\boldsymbol{\Theta}) : \ell^*(\boldsymbol{\theta}^*; \boldsymbol{S}) = \sup_{\boldsymbol{\theta} \in \boldsymbol{\Theta}} \ell^*(\boldsymbol{\theta}; \boldsymbol{S})\right\} \neq \emptyset\,.$$

*A11)* $P^*(\boldsymbol{\theta}) \leq 0$ *for all* $\boldsymbol{\theta} \in \boldsymbol{\Theta}$

*Then for any $\epsilon > 0$ there exists a $\delta > 0$ such that*

$$\|\boldsymbol{S} - \boldsymbol{\Sigma}_0\|_{\max} < \delta\,, \text{and } |n^{-1}P^*(\boldsymbol{\theta}_0)| < \delta \implies \left\|\boldsymbol{\Lambda}_0 - \boldsymbol{\Lambda}(\tilde{\boldsymbol{\theta}})\boldsymbol{Q}\right\|_{\max} < \epsilon\,, \left\|\boldsymbol{\Psi}_0 - \boldsymbol{\Psi}(\tilde{\boldsymbol{\theta}})\right\|_{\max} < \epsilon\,,$$

*for some orthogonal $q \times q$ matrix $\boldsymbol{Q}$.*

*Proof.* The proof follows the ideas of Kano (1983) whilst accommodating the penalty function $P^*(\boldsymbol{\theta})$. First, note that the MPL estimator of (4) can equivalently be defined as the minimiser

$$\tilde{\boldsymbol{\theta}} = \arg\min_{\boldsymbol{\theta} \in \boldsymbol{\Theta}} \left\{F(\boldsymbol{S}, \boldsymbol{\Sigma}(\boldsymbol{\theta})) - n^{-1}P^*(\boldsymbol{\theta})\right\}\,, \tag{S2}$$

where

$$F(\boldsymbol{\Sigma}_1, \boldsymbol{\Sigma}_2) = \frac{1}{2}\left\{\text{tr}\left(\boldsymbol{\Sigma}_2^{-1}\boldsymbol{\Sigma}_1\right) - p + \log\det(\boldsymbol{\Sigma}_2) - \log\det(\boldsymbol{\Sigma}_1)\right\}$$

is a criterion function that introduced in Kano (1983).

For notational convenience, let $\tilde{\boldsymbol{\Sigma}} = \boldsymbol{\Sigma}(\tilde{\boldsymbol{\theta}}) = \boldsymbol{\Lambda}(\tilde{\boldsymbol{\theta}})\boldsymbol{\Lambda}(\tilde{\boldsymbol{\theta}})^\top + \boldsymbol{\Psi}(\tilde{\boldsymbol{\theta}})$. Now by (S2), and since $P^*(\boldsymbol{\theta}) \leq 0$, $|n^{-1}P^*(\boldsymbol{\theta}_0)| < \delta$, it must hold that

$$F(\boldsymbol{S}, \tilde{\boldsymbol{\Sigma}}) \leq F(\boldsymbol{S}, \tilde{\boldsymbol{\Sigma}}) - n^{-1}P^*(\tilde{\boldsymbol{\theta}}) \leq F(\boldsymbol{S}, \boldsymbol{\Sigma}_0) - n^{-1}P^*(\boldsymbol{\theta}_0) \leq F(\boldsymbol{S}, \boldsymbol{\Sigma}_0) + \delta\,. \tag{S3}$$

Now for $\|\boldsymbol{S} - \boldsymbol{\Sigma}_0\|_F \leq \delta$ and $\delta$ small enough, where $\|\boldsymbol{A}\|_F = \text{tr}\left(\boldsymbol{A}^\top\boldsymbol{A}\right)^{1/2}$ is the Frobenius norm, it is shown in Kano (1983, equations A.2, A.4), that

$$F(\boldsymbol{S}, \boldsymbol{\Sigma}_0) \leq M\left\|\boldsymbol{\Sigma}_0^{-1/2}\right\|_F^4 \|\boldsymbol{\Sigma}_0 - \boldsymbol{S}\|_F^2 < M\left\|\boldsymbol{\Sigma}_0^{-1/2}\right\|_F^4 \delta^2\,,$$

and also

$$F(\boldsymbol{S}, \tilde{\boldsymbol{\Sigma}}) \geq m\left\|\tilde{\boldsymbol{\Sigma}}^{-1/2}\boldsymbol{S}\tilde{\boldsymbol{\Sigma}}^{-1/2} - \boldsymbol{I}_p\right\|_F^2 \geq m\left\|\boldsymbol{S} - \tilde{\boldsymbol{\Sigma}}\right\|_F^2 \left\|\tilde{\boldsymbol{\Sigma}}^{1/2}\right\|_F^{-4}\,. \tag{S4}$$

Hence, (S3)-(S4) yield

$$m\left\|\boldsymbol{S} - \tilde{\boldsymbol{\Sigma}}\right\|_F^2 \left\|\tilde{\boldsymbol{\Sigma}}^{1/2}\right\|_F^{-4} \leq m\left\|\tilde{\boldsymbol{\Sigma}}^{-1/2}\boldsymbol{S}\tilde{\boldsymbol{\Sigma}}^{-1/2} - \boldsymbol{I}_p\right\|_F^2 \leq \delta\left(1 + \delta M\left\|\boldsymbol{\Sigma}_0^{-1/2}\right\|_F^4\right)\,. \tag{S5}$$

It further holds by (S5) and for $\|\boldsymbol{S} - \boldsymbol{\Sigma}_0\|_{\max} \leq \delta$, $\delta$ sufficiently small, that $\left\|\tilde{\boldsymbol{\Sigma}}^{1/2}\right\|_F \leq C$ for some constant $C > 0$. Thus we conclude that

$$\left\|\boldsymbol{S} - \tilde{\boldsymbol{\Sigma}}\right\|_F^2 \leq \delta\left(1 + \delta M\left\|\boldsymbol{\Sigma}_0^{-1/2}\right\|_F^4\right)m^{-1}C^4\,,$$

which can be made arbitrarily small by choosing a small enough $\delta$. Hence,

$$\begin{aligned}\left\|\tilde{\boldsymbol{\Sigma}} - \boldsymbol{\Sigma}_0\right\|_{\max} &= \left\|\tilde{\boldsymbol{\Sigma}} - \boldsymbol{S} + (\boldsymbol{S} - \boldsymbol{\Sigma}_0)\right\|_{\max} \\ &\leq \left\|\tilde{\boldsymbol{\Sigma}} - \boldsymbol{S}\right\|_{\max} + \|\boldsymbol{S} - \boldsymbol{\Sigma}_0\|_{\max} \\ &\leq C'\delta + \delta\,,\end{aligned} \tag{S6}$$

where $C' > 0$ stems from (S5) and $\|\boldsymbol{A}\|_F = \sqrt{\sum_{i,j}|\boldsymbol{A}_{i,j}|^2} \geq \max_{i,j}\{|\boldsymbol{A}_{i,j}|\} = \|\boldsymbol{A}\|_{\max}$. Since the RHS in the last line of (S6) can be made arbitrarily small, the claim follows from $\tilde{\boldsymbol{\Sigma}} = \boldsymbol{\Sigma}(\tilde{\boldsymbol{\theta}}) = \boldsymbol{\Lambda}(\tilde{\boldsymbol{\theta}})\boldsymbol{\Lambda}(\tilde{\boldsymbol{\theta}})^\top + \boldsymbol{\Psi}(\tilde{\boldsymbol{\theta}})$, strong identifiability. $\square$

**Theorem A5.** *Assume that*

C1) *the factor model is strongly identifiable*

C2) *the set of maximum penalized likelihood estimates $\tilde{\boldsymbol{\theta}} \in \arg\max_{\boldsymbol{\theta} \in \Theta} \{\ell^*(\boldsymbol{\theta}; S)\}$ is non-empty; and*

C3) $P^*(\boldsymbol{\theta}) \leq 0$ *for all* $\boldsymbol{\theta} \in \Theta$

*Then, for any $\epsilon > 0$, there exists a $\delta > 0$ such that*

$$\|\boldsymbol{S} - \boldsymbol{\Sigma}_0\|_{\max} < \delta, \text{ and } |n^{-1} P^*(\boldsymbol{\theta}_0)| < \delta \implies \left\|\boldsymbol{\Lambda}_0 - \boldsymbol{\Lambda}(\tilde{\boldsymbol{\theta}})\boldsymbol{Q}\right\|_{\max} < \epsilon, \left\|\boldsymbol{\Psi}_0 - \boldsymbol{\Psi}(\tilde{\boldsymbol{\theta}})\right\|_{\max} < \epsilon,$$

*for some orthogonal $q \times q$ matrix $\boldsymbol{Q}$.*

*Proof.* Let $\boldsymbol{\Theta} = \{\boldsymbol{\theta} \in \Re^{p(q+1)} : \forall m > pq, \theta_m > 0\}$ and $\partial \boldsymbol{\Theta} = \{\boldsymbol{\theta} \in \Re^{p(q+1)} : \exists m > pq, \theta_m = 0\}$ be its boundary. The claim of the theorem immediately follows from Theorem A4. □

## S2.3 $\sqrt{n}$-consistency

We first state a blanket theorem for $\sqrt{n}$-consistency of MPL estimators, from which Theorem 3 follows.

**Theorem A6.** *Let*

$$\tilde{\boldsymbol{\theta}} \in \arg\max_{\boldsymbol{\theta} \in \Theta} \{\ell(\boldsymbol{\theta}) + P^*(\boldsymbol{\theta})\}, \quad \hat{\boldsymbol{\theta}} \in \arg\max_{\boldsymbol{\theta} \in \Theta} \{\ell(\boldsymbol{\theta})\}.$$

*Assume that conditions the conditions*

A12) *$\hat{\boldsymbol{\theta}}, \tilde{\boldsymbol{\theta}}$ exist and converge to $\boldsymbol{\theta}_0$, an interior point of $\boldsymbol{\Theta}$, with probability approaching one as $n \to \infty$*

A13) *In a convex neighbourhood $\mathcal{N}_0$ around $\boldsymbol{\theta}_0$, $\ell(\boldsymbol{\theta})$ is twice differentiable with gradient that is continuous on an open set containing the interior of $\mathcal{N}_0$*

A14) *$\sup_{\boldsymbol{\theta} \in \mathcal{N}_0} \left\|\boldsymbol{R}_n^{-1/2} \nabla \nabla^\top \ell(\boldsymbol{\theta}) \boldsymbol{R}_n^{-1/2} - \boldsymbol{J}(\boldsymbol{\theta})\right\| = o_p(1)$, where $\boldsymbol{J}(\boldsymbol{\theta})$ is deterministic, continuous and invertible at $\boldsymbol{\theta}_0$ and $\boldsymbol{R}_n$ is a sequence of diagonal, positive definite matrices indexed by $n$*

A15) *$P^*(\boldsymbol{\theta})$ is differentiable on $\mathcal{N}_0$ around $\boldsymbol{\theta}_0$, and $\sup_{\boldsymbol{\theta} \in \mathcal{N}_0} \|\boldsymbol{R}_n^{-1/2} \nabla P^*(\boldsymbol{\theta})\| = o_p(1)$*

*hold. Then*

$$\left\|\boldsymbol{R}_n^{1/2}(\hat{\boldsymbol{\theta}} - \tilde{\boldsymbol{\theta}})\right\| = o_p(1).$$

*Proof.* By the equivalence of norms on finite dimensional Euclidean spaces, without loss of generality, for the remainder of this proof, let $\|\boldsymbol{v}\| = \|\boldsymbol{v}\|_\infty = \sup_{1 \leq i \leq d} |\boldsymbol{v}_i|$ and $\|\boldsymbol{M}\| = \|\boldsymbol{M}\|_\infty$ be the corresponding operator norm.

Fix constants $\epsilon, \delta$. Define the events

$\mathcal{A}_n$ : $\hat{\boldsymbol{\theta}}, \tilde{\boldsymbol{\theta}} \in \text{int}(\boldsymbol{\Theta}) \cap \mathcal{N}_0 \cap B_\varepsilon(\boldsymbol{\theta}_0)$, where $\varepsilon$ is the constant for which by continuity of $\boldsymbol{J}(\boldsymbol{\theta}_0)$, it holds that $\|\boldsymbol{\theta} - \boldsymbol{\theta}_0\| < \varepsilon$ implies $\|\boldsymbol{J}(\boldsymbol{\theta}) - \boldsymbol{J}(\boldsymbol{\theta}_0)\| < \{4\|\boldsymbol{J}(\boldsymbol{\theta}_0)^{-1}\|\}^{-1}$ and $B_\varepsilon(\boldsymbol{\theta}_0) = \{\boldsymbol{\theta} \in \boldsymbol{\Theta} : \|\boldsymbol{\theta} - \boldsymbol{\theta}_0\| < \varepsilon\}$.

$\mathcal{B}_n$ : $\sup_{\boldsymbol{\theta} \in \mathcal{N}_0} \|\boldsymbol{H}_n(\boldsymbol{\theta}) - \boldsymbol{J}(\boldsymbol{\theta})\| \leq \{4\|\boldsymbol{J}(\boldsymbol{\theta}_0)^{-1}\|\}^{-1}$, where $\boldsymbol{H}_n = \boldsymbol{R}_n^{-1/2} \nabla \nabla^\top \ell(\boldsymbol{\theta}) \boldsymbol{R}_n^{-1/2}$

$\mathcal{C}_n$ : $P^*(\boldsymbol{\theta})$ is differentiable in $\mathcal{N}_0$ and $\sup_{\boldsymbol{\theta} \in \mathcal{N}_0} \|\boldsymbol{R}_n^{-1/2} \nabla P^*(\boldsymbol{\theta})\| \leq \epsilon/2\|\boldsymbol{J}(\boldsymbol{\theta}_0)^{-1}\|$.

Assume that $\mathcal{A}_n \cap \mathcal{B}_n \cap \mathcal{C}_n$ holds. Then by $\mathcal{A}_n$, and assumptions A13) and A15),

$$\mathbf{0} = \nabla \ell(\hat{\boldsymbol{\theta}})$$
$$\mathbf{0} = \nabla \ell(\tilde{\boldsymbol{\theta}}) + \nabla P^*(\tilde{\boldsymbol{\theta}}),$$

where $\nabla \ell(\boldsymbol{\theta})$ denotes the gradient of $\ell(\boldsymbol{\theta})$ with respect to $\boldsymbol{\theta}$. Thus,

$$\nabla \ell(\hat{\boldsymbol{\theta}}) - \nabla \ell(\tilde{\boldsymbol{\theta}}) = \nabla P^*(\tilde{\boldsymbol{\theta}}).$$

By the Mean Value Theorem (see, for example, Rudin, 1976, Theorem 5.10), the $i$th component of the equation above can be written as

$$\left\{\nabla \nabla^\top \ell(\boldsymbol{\theta}_i^*)\right\}_{i,\bullet}^\top (\hat{\boldsymbol{\theta}} - \tilde{\boldsymbol{\theta}}) = \left\{\nabla P^*(\tilde{\boldsymbol{\theta}})\right\}_i,$$

where $\{\boldsymbol{A}\}_{i,\bullet}$ denotes the $i$th row of matrix $\boldsymbol{A}$ and $\boldsymbol{\theta}_i^*$ is a vector on the line segment joining $\hat{\boldsymbol{\theta}}$ and $\tilde{\boldsymbol{\theta}}$, i.e. $\boldsymbol{\theta}_i^* = c_i \hat{\boldsymbol{\theta}} + (1 - c_i) \tilde{\boldsymbol{\theta}}$, $c_i \in [0, 1]$. Abusing notation, let $\nabla \nabla^\top \ell(\boldsymbol{\theta}^*)$ be the matrix with rows $\{\nabla \nabla^\top \ell(\boldsymbol{\theta}_i^*)\}_{i,\bullet}$ and let $\boldsymbol{H}_n = \boldsymbol{R}_n^{-1/2} \nabla \nabla^\top \ell(\boldsymbol{\theta}^*) \boldsymbol{R}_n^{-1/2}$. Then

$$\boldsymbol{J}(\boldsymbol{\theta}_0)^{-1} \boldsymbol{H}_n \boldsymbol{R}_n^{1/2}(\hat{\boldsymbol{\theta}} - \tilde{\boldsymbol{\theta}}) = \boldsymbol{J}(\boldsymbol{\theta}_0)^{-1} \boldsymbol{R}_n^{-1/2} \nabla P^*(\tilde{\boldsymbol{\theta}}),$$

or equivalently

$$\boldsymbol{J}(\boldsymbol{\theta}_0)^{-1} \left\{\boldsymbol{H}_n - \boldsymbol{J}(\boldsymbol{\theta}_0)\right\} \boldsymbol{R}_n^{1/2}(\hat{\boldsymbol{\theta}} - \tilde{\boldsymbol{\theta}}) + \boldsymbol{R}_n^{1/2}(\hat{\boldsymbol{\theta}} - \tilde{\boldsymbol{\theta}}) = \boldsymbol{J}(\boldsymbol{\theta}_0)^{-1} \boldsymbol{R}_n^{-1/2} \nabla P^*(\tilde{\boldsymbol{\theta}}),$$

and rearranging yields

$$\boldsymbol{R}_n^{1/2}(\hat{\boldsymbol{\theta}} - \tilde{\boldsymbol{\theta}}) = \boldsymbol{J}(\boldsymbol{\theta}_0)^{-1} \boldsymbol{R}_n^{-1/2} \nabla P^*(\tilde{\boldsymbol{\theta}}) + \boldsymbol{J}(\boldsymbol{\theta}_0)^{-1} \left\{\boldsymbol{J}(\boldsymbol{\theta}_0) - \boldsymbol{H}_n\right\} \boldsymbol{R}_n^{1/2}(\hat{\boldsymbol{\theta}} - \tilde{\boldsymbol{\theta}}).$$

Then

$$\left\| \boldsymbol{R}_n^{1/2}(\hat{\boldsymbol{\theta}} - \tilde{\boldsymbol{\theta}}) \right\| = \left\| \boldsymbol{J}(\boldsymbol{\theta}_0)^{-1} \boldsymbol{R}_n^{-1/2} \nabla P^*(\tilde{\boldsymbol{\theta}}) + \boldsymbol{J}(\boldsymbol{\theta}_0)^{-1} \left\{\boldsymbol{J}(\boldsymbol{\theta}_0) - \boldsymbol{H}_n\right\} \boldsymbol{R}_n^{1/2}(\hat{\boldsymbol{\theta}} - \tilde{\boldsymbol{\theta}}) \right\|$$
$$\leq \left\|\left\| \boldsymbol{J}(\boldsymbol{\theta}_0)^{-1} \right\|\right\| \left\| \boldsymbol{R}_n^{-1/2} \nabla P^*(\tilde{\boldsymbol{\theta}}) \right\| + \left\|\left\| \boldsymbol{J}(\boldsymbol{\theta}_0)^{-1} \right\|\right\| \left\|\left\| \boldsymbol{J}(\boldsymbol{\theta}_0) - \boldsymbol{H}_n \right\|\right\| \left\| \boldsymbol{R}_n^{1/2}(\hat{\boldsymbol{\theta}} - \tilde{\boldsymbol{\theta}}) \right\|. \quad (S7)$$

Now let $i$ be the row for which the row sum of $\boldsymbol{J}(\boldsymbol{\theta}_0) - \boldsymbol{H}_n$ is maximal, then by $\mathcal{A}_n, \mathcal{B}_n$,

$$\begin{aligned}
\left\|\left\| \boldsymbol{J}(\boldsymbol{\theta}_0) - \boldsymbol{H}_n \right\|\right\| &= \sum_{j=1}^{p} |\boldsymbol{J}(\boldsymbol{\theta}_0)_{ij} - \boldsymbol{H}_n(\boldsymbol{\theta}_i^*)_{ij}| \\
&= \sum_{j=1}^{p} |\boldsymbol{J}(\boldsymbol{\theta}_0)_{ij} - \boldsymbol{J}(\boldsymbol{\theta}_i^*)_{ij} + \boldsymbol{J}(\boldsymbol{\theta}_i^*)_{ij} - \boldsymbol{H}_n(\boldsymbol{\theta}_i^*)_{ij}| \\
&\leq \sum_{j=1}^{p} |\boldsymbol{J}(\boldsymbol{\theta}_0)_{ij} - \boldsymbol{J}(\boldsymbol{\theta}_i^*)_{ij}| + \sum_{j=1}^{p} |\boldsymbol{J}(\boldsymbol{\theta}_i^*)_{ij} - \boldsymbol{H}_n(\boldsymbol{\theta}_i^*)_{ij}| \\
&\leq \sup_{1 \leq k \leq p} \sum_{j=1}^{p} |\boldsymbol{J}(\boldsymbol{\theta}_0)_{kj} - \boldsymbol{J}(\boldsymbol{\theta}_i^*)_{kj}| + \sup_{1 \leq k \leq p} \sum_{j=1}^{p} |\boldsymbol{J}(\boldsymbol{\theta}_i^*)_{kj} - \boldsymbol{H}_n(\boldsymbol{\theta}_i^*)_{kj}| \\
&= \left\|\left\| \boldsymbol{J}(\boldsymbol{\theta}_0) - \boldsymbol{J}(\boldsymbol{\theta}_i^*) \right\|\right\| + \left\|\left\| \boldsymbol{J}(\boldsymbol{\theta}_i^*) - \boldsymbol{H}_n(\boldsymbol{\theta}_i^*) \right\|\right\| \\
&\leq \left\|\left\| \boldsymbol{J}(\boldsymbol{\theta}_0) - \boldsymbol{J}(\boldsymbol{\theta}_i^*) \right\|\right\| + \sup_{\boldsymbol{\theta} \in \mathcal{N}_0} \left\|\left\| \boldsymbol{J}(\boldsymbol{\theta}) - \boldsymbol{H}_n(\boldsymbol{\theta}) \right\|\right\| \\
&\leq \frac{1}{4 \left\|\left\| \boldsymbol{J}(\boldsymbol{\theta}_0)^{-1} \right\|\right\|} + \frac{1}{4 \left\|\left\| \boldsymbol{J}(\boldsymbol{\theta}_0)^{-1} \right\|\right\|} \\
&= \frac{1}{2 \left\|\left\| \boldsymbol{J}(\boldsymbol{\theta}_0)^{-1} \right\|\right\|},
\end{aligned} \quad (S8)$$

11

where we used that $\mathcal{A}_n$ and that $\boldsymbol{\theta}_i = c_i\hat{\boldsymbol{\theta}} + (1 - c_i)\tilde{\boldsymbol{\theta}}$ so $\|\boldsymbol{\theta}_i^* - \boldsymbol{\theta}_0\| \leq c_i\|\hat{\boldsymbol{\theta}} - \boldsymbol{\theta}_0\| + (1 - c_i)\|\tilde{\boldsymbol{\theta}} - \boldsymbol{\theta}_0\| < \varepsilon$.

Hence, substituting (S8) into (S7), upon rearranging, yields

$$\left\|\boldsymbol{R}_n^{1/2}(\hat{\boldsymbol{\theta}} - \tilde{\boldsymbol{\theta}})\right\| \leq 2\||\boldsymbol{J}(\boldsymbol{\theta}_0)^{-1}\|| \left\|\boldsymbol{R}_n^{-1/2}\nabla P^*(\tilde{\boldsymbol{\theta}})\right\|,$$

and by $\mathcal{C}_n$, we conclude that

$$\begin{aligned}
\left\|\boldsymbol{R}_n^{1/2}(\hat{\boldsymbol{\theta}} - \tilde{\boldsymbol{\theta}})\right\| &\leq 2\||\boldsymbol{J}(\boldsymbol{\theta}_0)^{-1}\|| \left\|\boldsymbol{R}_n^{-1/2}\nabla P^*(\tilde{\boldsymbol{\theta}})\right\| \\
&\leq 2\||\boldsymbol{J}(\boldsymbol{\theta}_0)^{-1}\|| \sup_{\boldsymbol{\theta} \in \mathcal{N}_0} \left\|\boldsymbol{R}_n^{-1/2}\nabla P^*(\boldsymbol{\theta})\right\| \\
&\leq 2\||\boldsymbol{J}(\boldsymbol{\theta}_0)^{-1}\|| \frac{\epsilon}{2\||\boldsymbol{J}(\boldsymbol{\theta}_0)^{-1}\||} \\
&= \epsilon.
\end{aligned}$$

Hence, we have shown that $\mathcal{A}_n \cap \mathcal{B}_n \cap \mathcal{C}_n$ implies $\|\boldsymbol{R}_n^{1/2}(\hat{\boldsymbol{\theta}} - \tilde{\boldsymbol{\theta}})\| \leq \epsilon$. Therefore,

$$\Pr\left(\left\|\boldsymbol{R}_n^{1/2}(\hat{\boldsymbol{\theta}} - \tilde{\boldsymbol{\theta}})\right\| > \epsilon\right) \leq 1 - \Pr(\mathcal{A}_n \cap \mathcal{B}_n \cap \mathcal{C}_n).$$

By assumptions A12)-A15), there exists a $N$ such that for all $n > N$,

$$\begin{aligned}
\Pr(\neg\mathcal{A}_n) &\leq \frac{\delta}{3} \\
\Pr(\neg\mathcal{B}_n) &\leq \frac{\delta}{3} \\
\Pr(\neg\mathcal{C}_n) &\leq \frac{\delta}{3},
\end{aligned}$$

And thus by a union bound,

$$\Pr(\mathcal{A}_n \cap \mathcal{B}_n \cap \mathcal{C}_n) = 1 - \Pr(\neg\mathcal{A}_n \cup \neg\mathcal{B}_n \cup \neg\mathcal{C}_n) \geq 1 - \delta,$$

so that we conclude that for any $\delta, \epsilon$, there is a $N$ such that for all $n > N$

$$\Pr\left(\left\|\boldsymbol{R}_n^{1/2}(\hat{\boldsymbol{\theta}} - \tilde{\boldsymbol{\theta}})\right\| > \epsilon\right) \leq \delta.$$

$\square$

**Theorem A7.** *Suppose that*

N1) *there exists an interior point $\boldsymbol{\theta}_0 \in \boldsymbol{\Theta}$ such that $\boldsymbol{S} \xrightarrow{p} \boldsymbol{\Sigma}(\boldsymbol{\theta}_0)$ as $n \to \infty$;*

N2) *the factor model is strongly identifiable and the Jacobian of $vec(\boldsymbol{\Sigma}(\boldsymbol{\theta}))$ with respect to $\boldsymbol{\theta}$ is nonsingular at $\boldsymbol{\theta}_0$;*

N3) *the criterion function*

$$F(\boldsymbol{\Sigma}_1, \boldsymbol{\Sigma}_2) = \log\det(\boldsymbol{\Sigma}_2) + tr\left(\boldsymbol{\Sigma}_2^{-1}\boldsymbol{\Sigma}_1\right) - p - \log\det(\boldsymbol{\Sigma}_1)$$

*satisfies that for any $\epsilon > 0$, there exists a $\delta > 0$ such that if $\||\boldsymbol{S} - \boldsymbol{\Sigma}_0\|| < \delta$ and $F(\boldsymbol{S}, \boldsymbol{\Sigma}(\boldsymbol{\theta})) < \delta$, then $\||\boldsymbol{\Sigma}(\boldsymbol{\theta}) - \boldsymbol{\Sigma}_0\|| < \epsilon$.*

N4) *the set of maximum penalized likelihood estimates $\tilde{\boldsymbol{\theta}} \in \arg\max_{\boldsymbol{\theta} \in \boldsymbol{\Theta}} \{\ell^*(\boldsymbol{\theta}; S)\}$ is not empty; and*

N5) *$P^*(\boldsymbol{\theta}) \leq 0$, and $P^*(\boldsymbol{\theta}) = o_p(\sqrt{n})$ where $P(\boldsymbol{\theta})$ is continuously differentiable on $\boldsymbol{\Theta}$ and invariant under orthogonal rotations of $\boldsymbol{\Lambda}$*

*Then, there exist sequences of orthogonal rotation matrices $\boldsymbol{Q}_1, \boldsymbol{Q}_2$ such that:*

$$\left\|\boldsymbol{\Lambda}(\tilde{\boldsymbol{\theta}})\boldsymbol{Q}_1 - \boldsymbol{\Lambda}_0\right\| \xrightarrow{p} 0, \quad \left\|\boldsymbol{\Psi}(\tilde{\boldsymbol{\theta}}) - \boldsymbol{\Psi}_0\right\| \xrightarrow{p} 0,$$

*and*

$$\sqrt{n}\left\|\boldsymbol{\Lambda}(\tilde{\boldsymbol{\theta}})\boldsymbol{Q}_1 - \boldsymbol{\Lambda}(\hat{\boldsymbol{\theta}})\boldsymbol{Q}_2\right\| \xrightarrow{p} 0, \quad \sqrt{n}\left\|\boldsymbol{\Psi}(\tilde{\boldsymbol{\theta}}) - \boldsymbol{\Psi}(\hat{\boldsymbol{\theta}})\right\| \xrightarrow{p} 0.$$

*Proof.* We shall prove that the conditions of Theorem A6 are met.

A12) Let $\hat{\boldsymbol{\Lambda}}, \hat{\boldsymbol{\Psi}}$ be $\boldsymbol{\Lambda}(\hat{\boldsymbol{\theta}}), \boldsymbol{\Psi}(\hat{\boldsymbol{\theta}})$, be the loading matrix and the variances of the ML estimator $\hat{\boldsymbol{\theta}}$, respectively. Lemma 1 shows that there exists a sequence of orthogonal rotation matrices $\boldsymbol{Q}_1$, such that $\{\text{vec}(\hat{\boldsymbol{\Lambda}}\boldsymbol{Q}_1), \text{diag}(\hat{\boldsymbol{\Psi}})\} \xrightarrow{p} \boldsymbol{\theta}_0 = (\text{vec}(\boldsymbol{\Lambda}_0)^\top, \text{diag}(\boldsymbol{\Psi}_0)^\top)^\top$. Since this vector is also a maximiser of the log-likelihood, we henceforth assume that $\hat{\boldsymbol{\theta}}$ is chosen with the adequate rotation $\boldsymbol{Q}_1$ such that $\hat{\boldsymbol{\theta}} \xrightarrow{p} \boldsymbol{\theta}_0$.

Similarly, for the MPL estimator $\tilde{\boldsymbol{\theta}}$, note that assumptions N1)-N5) satisfy the conditions for consistency in Theorem A4. Hence, there exists a sequence of orthogonal rotation matrices $\boldsymbol{Q}_2$, such that $\{\text{vec}(\tilde{\boldsymbol{\Lambda}}\boldsymbol{Q}_2), \text{diag}(\tilde{\boldsymbol{\Psi}})\} \xrightarrow{p} \boldsymbol{\theta}_0$. By assumption N5), $P(\boldsymbol{\theta})$ is invariant under such orthogonal rotations, so that $\{\text{vec}(\tilde{\boldsymbol{\Lambda}}\boldsymbol{Q}_2), \text{diag}(\tilde{\boldsymbol{\Psi}})\}$ is also a maximiser of the penalised log-likelihood. Thus, henceforth assume that $\tilde{\boldsymbol{\theta}}$ chooses the adequate rotation $\boldsymbol{Q}_2$ such that $\tilde{\boldsymbol{\theta}} \xrightarrow{p} \boldsymbol{\theta}_0$.

A13) Twice-Differentiability follows by twice-differentiability of $\ell(\boldsymbol{\Sigma}; \boldsymbol{S})$ with respect to $\boldsymbol{\Sigma}$ and the construction of $\boldsymbol{\Sigma}$. Note that the partial derivatives are given by

$$\frac{\partial \ell(\boldsymbol{\Sigma}(\boldsymbol{\theta}); \boldsymbol{S})}{\partial \boldsymbol{\theta}_j} = -\frac{n}{2}\left[\text{tr}\left\{\boldsymbol{\Sigma}(\boldsymbol{\theta})^{-1}\frac{\partial \boldsymbol{\Sigma}(\boldsymbol{\theta})}{\partial \boldsymbol{\theta}_j}\right\} - \text{tr}\left\{\boldsymbol{\Sigma}(\boldsymbol{\theta})^{-1}\frac{\partial \boldsymbol{\Sigma}(\boldsymbol{\theta})}{\partial \boldsymbol{\theta}_j}\boldsymbol{\Sigma}(\boldsymbol{\theta})^{-1}\boldsymbol{S}\right\}\right]$$

which are continuous in $\boldsymbol{\Sigma}$ and $\boldsymbol{\Sigma}(\boldsymbol{\theta})$ is continuous in $\boldsymbol{\theta}$ on $\boldsymbol{\Theta}$ by N2).

A14) Note that

$$\frac{\partial^2 \ell(\boldsymbol{\Sigma}(\boldsymbol{\theta}); \boldsymbol{S})}{\partial \boldsymbol{\theta}_j \partial \boldsymbol{\theta}_k} = -\frac{n}{2}\left[-\text{tr}\left\{\boldsymbol{\Sigma}(\boldsymbol{\theta})^{-1}\frac{\partial \boldsymbol{\Sigma}(\boldsymbol{\theta})}{\partial \boldsymbol{\theta}_j}\boldsymbol{\Sigma}(\boldsymbol{\theta})^{-1}\frac{\partial \boldsymbol{\Sigma}(\boldsymbol{\theta})}{\partial \boldsymbol{\theta}_i}\right\}\right.$$
$$+ \text{tr}\left\{\boldsymbol{\Sigma}(\boldsymbol{\theta})^{-1}\frac{\partial^2 \boldsymbol{\Sigma}(\boldsymbol{\theta})}{\partial \boldsymbol{\theta}_i \partial \boldsymbol{\theta}_j}\right\}$$
$$+ 2\,\text{tr}\left\{\boldsymbol{\Sigma}(\boldsymbol{\theta})^{-1}\frac{\partial \boldsymbol{\Sigma}(\boldsymbol{\theta})}{\partial \boldsymbol{\theta}_i}\boldsymbol{\Sigma}(\boldsymbol{\theta})^{-1}\frac{\partial \boldsymbol{\Sigma}(\boldsymbol{\theta})}{\partial \boldsymbol{\theta}_j}\boldsymbol{\Sigma}(\boldsymbol{\theta})^{-1}\boldsymbol{S}\right\}$$
$$\left. - \text{tr}\left\{\boldsymbol{\Sigma}(\boldsymbol{\theta})^{-1}\frac{\partial^2 \boldsymbol{\Sigma}(\boldsymbol{\theta})}{\partial \boldsymbol{\theta}_i \partial \boldsymbol{\theta}_j}\boldsymbol{\Sigma}(\boldsymbol{\theta})^{-1}\boldsymbol{S}\right\}\right]$$

and let the $j, k$th entry of $\boldsymbol{J}(\boldsymbol{\theta})$ be given by $n^{-1}\partial^2\ell(\boldsymbol{\Sigma}(\boldsymbol{\theta}); \boldsymbol{\Sigma}_0)/\partial\boldsymbol{\theta}_j\partial\boldsymbol{\theta}_j$, which is clearly continuous in $\boldsymbol{\theta}$. Let $\mathcal{N}_0$ be an Euclidean ball that lies fully in the interior of $\boldsymbol{\Theta}$ and which is centred around $\boldsymbol{\theta}_0$. Finally let $\boldsymbol{R}_n = n\boldsymbol{I}_p$. Then

$$\sup_{\boldsymbol{\theta} \in \mathcal{N}_0}\left|n^{-1}\frac{\partial^2 \ell(\boldsymbol{\Sigma}(\boldsymbol{\theta}); \boldsymbol{S})}{\partial \boldsymbol{\theta}_j \partial \boldsymbol{\theta}_k} - \boldsymbol{J}(\boldsymbol{\theta})_{ij}\right| = \sup_{\boldsymbol{\theta} \in \mathcal{N}_0}|\text{tr}\{\boldsymbol{B}(\boldsymbol{\theta})[\boldsymbol{S} - \boldsymbol{\Sigma}_0]\}|$$
$$\leq \sup_{\boldsymbol{\theta} \in \mathcal{N}_0}\|\|\boldsymbol{B}(\boldsymbol{\theta})\|\|_*\|\|\boldsymbol{S} - \boldsymbol{\Sigma}_0\|\|_2$$
$$\leq C\|\|\boldsymbol{S} - \boldsymbol{\Sigma}_0\|\|_2$$
$$= o_p(1),$$

where the second line follows by Hoelder's inequality for Schatten-norms and where $\|\|\boldsymbol{A}\|\|_*$ is the sum of all singular values of $\boldsymbol{A}$ and $\|\|\boldsymbol{A}\|\|_2$ is the spectral norm. The third line follows by the Extreme Value Theorem (see, for example, Rudin, 1976, Theorem 4.16) since $\|\|\boldsymbol{B}(\boldsymbol{\theta})\|\|_*$, which is

$$\boldsymbol{B}(\boldsymbol{\theta}) = \text{tr}\left\{\boldsymbol{\Sigma}(\boldsymbol{\theta})^{-1}\frac{\partial \boldsymbol{\Sigma}(\boldsymbol{\theta})}{\partial \boldsymbol{\theta}_i}\boldsymbol{\Sigma}(\boldsymbol{\theta})^{-1}\frac{\partial \boldsymbol{\Sigma}(\boldsymbol{\theta})}{\partial \boldsymbol{\theta}_j}\boldsymbol{\Sigma}(\boldsymbol{\theta})^{-1} - \frac{1}{2}\boldsymbol{\Sigma}(\boldsymbol{\theta})^{-1}\frac{\partial^2 \boldsymbol{\Sigma}(\boldsymbol{\theta})}{\partial \boldsymbol{\theta}_i \partial \boldsymbol{\theta}_j}\boldsymbol{\Sigma}(\boldsymbol{\theta})^{-1}\right\},$$

13

is continuous in $\boldsymbol{\theta}$ and $\mathcal{N}_0$ is compact. The last line follows since $\boldsymbol{S}$ converges to $\boldsymbol{\Sigma}_0$ in probability as $n \to \infty$ (N1)). Finally, to see that $\boldsymbol{J}(\boldsymbol{\theta}_0)$ is invertible, note that

$$\boldsymbol{J}(\boldsymbol{\theta}_0)_{ij} = \frac{1}{2} \operatorname{tr} \left\{ \boldsymbol{\Sigma}_0^{-1} \frac{\partial \boldsymbol{\Sigma}(\boldsymbol{\theta}_0)}{\partial \boldsymbol{\theta}_j} \boldsymbol{\Sigma}_0^{-1} \frac{\partial \boldsymbol{\Sigma}(\boldsymbol{\theta}_0)}{\partial \boldsymbol{\theta}_i} \right\}.$$

Hence, we can write $\boldsymbol{J}(\boldsymbol{\theta}_0)$ as

$$\boldsymbol{J}(\boldsymbol{\theta}_0) = -\frac{1}{2} \mathcal{J}^\top \left( \boldsymbol{\Sigma}_0^{-1} \otimes \boldsymbol{\Sigma}_0^{-1} \right) \mathcal{J},$$

where

$$\mathcal{J} = \left[ \operatorname{vec}\left(\frac{\partial \boldsymbol{\Sigma}(\boldsymbol{\theta}_0)}{\partial \boldsymbol{\theta}_1}\right), \quad \operatorname{vec}\left(\frac{\partial \boldsymbol{\Sigma}(\boldsymbol{\theta}_0)}{\partial \boldsymbol{\theta}_2}\right), \quad \cdots, \quad \operatorname{vec}\left(\frac{\partial \boldsymbol{\Sigma}(\boldsymbol{\theta}_0)}{\partial \boldsymbol{\theta}_d}\right) . \right]$$

Hence $\boldsymbol{J}(\boldsymbol{\theta}_0)$ is invertible if $\mathcal{J}$ has full column rank. This holds by our identification condition N2).

A15) Let $\mathcal{N}_0$ be as defined above. By assumption N5), $\nabla P^*(\boldsymbol{\theta})$ is continuous and hence, by the Extreme Value Theorem,

$$\sup_{\boldsymbol{\theta} \in \mathcal{N}} \left\| n^{-1/2} \nabla P^*(\boldsymbol{\theta}) \right\| = \frac{c_n}{\sqrt{n}} \sup_{\boldsymbol{\theta} \in \mathcal{N}} \|\nabla P(\boldsymbol{\theta})\|$$
$$\leq C \frac{c_n}{\sqrt{n}}$$
$$= o_p(1),$$

as required.

$\square$

**Lemma 1.** *In the setting of Theorem 3 with assumptions N1)-N3), with probability going to one, the ML estimator exists, and there exists a sequence of orthogonal rotation matrices $\boldsymbol{Q}$ such that*

$$\left\| \boldsymbol{\Lambda}(\hat{\boldsymbol{\theta}}) \boldsymbol{Q} - \boldsymbol{\Lambda}_0 \right\| \xrightarrow{p} 0, \quad \left\| \boldsymbol{\Psi}(\hat{\boldsymbol{\theta}}) - \boldsymbol{\Psi}_0 \right\| \xrightarrow{p} 0.$$

*Proof.* Consider the parameter space $\boldsymbol{\Theta}^* = \operatorname{vec}(\operatorname{vec}(\boldsymbol{\Theta}_1) \otimes \Re_{>0}^p)$, where $\boldsymbol{\Theta}_1$ is the space of all symmetric, positive semidefinite $p \times p$ matrices with rank $q$. For $\boldsymbol{\theta} \in \boldsymbol{\Theta}^*$ construct the variance-covariance matrix $\boldsymbol{s}(\boldsymbol{\theta})$ by stacking the first $p^2$ elements of $\boldsymbol{\theta}$ into the symmetric, positive semidefinite matrix $\boldsymbol{M}$ and the remaining $p$ entries of $\boldsymbol{\theta}$ into the diagonal matrix $\boldsymbol{U}$ and let $\boldsymbol{s}(\boldsymbol{\theta}) = \boldsymbol{M}(\boldsymbol{\theta}) + \boldsymbol{U}(\boldsymbol{\theta})$.

Kano (1986, Theorems 3, 4) shows that if $\boldsymbol{\Sigma}_0$ is strongly identifiable in $\boldsymbol{\Theta}^*$ and the criterion function, which we take to be the criterion function of N3) $\boldsymbol{\theta} \in \boldsymbol{\Theta}^*$, i.e.

$$F(\boldsymbol{\Sigma}, \boldsymbol{s}(\boldsymbol{\theta})) = \log \det(\boldsymbol{s}(\boldsymbol{\theta})) + \operatorname{tr}\left( (\boldsymbol{s}(\boldsymbol{\theta}))^{-1} \boldsymbol{\Sigma} \right) - p - \log \det(\boldsymbol{\Sigma}),$$

satisfies the condition (A2) (Kano, 1986, Section 3):

(A2) *For any $\epsilon$, there is a scalar $\delta > 0$ such that $\|\boldsymbol{S} - \boldsymbol{\Sigma}_0\| < \delta$ and $F(\boldsymbol{S}, \boldsymbol{s}(\boldsymbol{\theta})) < \delta$ imply that $\|\boldsymbol{s}(\boldsymbol{\theta}) - \boldsymbol{\Sigma}_0\| < \epsilon$,*

then with probability approaching to one, the minimiser of $F(\boldsymbol{S}, \boldsymbol{s}(\boldsymbol{\theta}))$ over $\boldsymbol{\Theta}^*$, exists and is consistent for $\boldsymbol{\Lambda}_0 \boldsymbol{\Lambda}_0^\top, \boldsymbol{\Psi}_0$, i.e.

$$\left\| \boldsymbol{M}(\hat{\boldsymbol{\theta}}) - \boldsymbol{\Lambda}_0 \boldsymbol{\Lambda}_0^\top \right\| \xrightarrow{p} 0, \quad \left\| \boldsymbol{U}(\hat{\boldsymbol{\theta}}) - \boldsymbol{\Psi}_0 \right\| \xrightarrow{p} 0. \tag{S9}$$

Hence, we need to show that

i) Any maximiser in $\boldsymbol{\Theta}^*$ is equivalent to a maximiser in $\boldsymbol{\Theta}$ in that $M(\hat{\boldsymbol{\theta}}) = \boldsymbol{\Lambda}(\hat{\boldsymbol{\theta}}) \boldsymbol{\Lambda}(\hat{\boldsymbol{\theta}})$ for some $p \times q$ matrix $\boldsymbol{\Lambda}(\hat{\boldsymbol{\theta}})$ of rank $q$, that, in conjunction with $\boldsymbol{U}(\hat{\boldsymbol{\theta}})$ is a maximiser of the log-likelihood in $\boldsymbol{\Theta}$

ii) Strong identifiability in $\boldsymbol{\Theta}$ implies strong identifiability in $\boldsymbol{\Theta}^*$,

iii) condition (A2) holds, and

14

iv) (S9) implies that there exists a sequence of orthogonal rotation matrices $\boldsymbol{Q}$ such that $\boldsymbol{\Lambda}(\hat{\boldsymbol{\theta}})\boldsymbol{Q} \to \boldsymbol{\Lambda}_0$ in probability.

We shall prove each property in turn.

i) Note that the parameter spaces $\boldsymbol{\Theta}$ and $\boldsymbol{\Theta}^*$ have the same images under the maps $\boldsymbol{\theta} \mapsto \boldsymbol{\Sigma}(\boldsymbol{\theta}), (\boldsymbol{\theta} \in \boldsymbol{\Theta})$, $\boldsymbol{\theta} \mapsto \boldsymbol{s}(\boldsymbol{\theta}), (\boldsymbol{\theta} \in \boldsymbol{\Theta}^*)$ respectively, as any $\boldsymbol{\Lambda}\boldsymbol{\Lambda}^\top$ is a symmetric, positive semi-definite $p \times p$ matrix, and conversely any such matrix $\boldsymbol{M}$ admits a representation $\boldsymbol{M} = \boldsymbol{\Lambda}\boldsymbol{\Lambda}^\top$, where $\boldsymbol{\Lambda}$ is $p \times q$ with rank $q$. The property follows by noting that minimising the criterion function $F$ is equivalent to maximising the log-likelihood function.

ii) If $\boldsymbol{\Sigma}_0 = \boldsymbol{\Lambda}_0\boldsymbol{\Lambda}_0^\top + \boldsymbol{\Psi}_0$ is strongly identifiable in $\boldsymbol{\Theta}$, then it is also strongly identifiable in $\boldsymbol{\Theta}^*$ in that for any $\epsilon > 0$, there is a $\delta > 0$ such that for $\boldsymbol{s} = \boldsymbol{M} + \boldsymbol{U}$, $\boldsymbol{M}$ symmetric positive definite $p \times p$ of rank $q$ and $\boldsymbol{U}$ $p \times p$ diagonal positive definite,

$$\|\|\boldsymbol{s} - \boldsymbol{\Sigma}_0\|\| < \delta \implies \|\|\boldsymbol{M} - \boldsymbol{\Lambda}_0\boldsymbol{\Lambda}_0^\top\|\| < \epsilon, \quad \|\|\boldsymbol{U} - \boldsymbol{\Psi}_0\|\| < \epsilon.$$

To see this write $\boldsymbol{s} = \boldsymbol{M} + \boldsymbol{U}$ as $\boldsymbol{s} = \boldsymbol{\Lambda}\boldsymbol{\Lambda}^\top + \boldsymbol{U}$. By strong identifiability of $\boldsymbol{\Sigma}_0$ in $\boldsymbol{\Theta}$, we have that

$$\|\|\boldsymbol{\Lambda}\boldsymbol{Q} - \boldsymbol{\Lambda}_0\|\| < \epsilon, \quad \|\|\boldsymbol{U} - \boldsymbol{\Psi}_0\|\| < \epsilon.$$

Now let $\boldsymbol{\Delta} = \boldsymbol{\Lambda}\boldsymbol{Q} - \boldsymbol{\Lambda}_0$. Then

$$\begin{aligned}
\|\|\boldsymbol{\Lambda}\boldsymbol{\Lambda}^\top - \boldsymbol{\Lambda}_0\boldsymbol{\Lambda}_0^\top\|\| &= \|\|(\boldsymbol{\Lambda}_0 - \boldsymbol{\Delta})(\boldsymbol{\Lambda}_0 - \boldsymbol{\Delta})^\top - \boldsymbol{\Lambda}_0\boldsymbol{\Lambda}_0^\top\|\| \\
&= \|\|\boldsymbol{\Lambda}_0\boldsymbol{\Delta}^\top + \boldsymbol{\Delta}\boldsymbol{\Lambda}_0^\top + \boldsymbol{\Delta}\boldsymbol{\Delta}^\top\|\| \\
&\leq 2\|\|\boldsymbol{\Lambda}_0\boldsymbol{\Delta}^\top\|\| + \|\|\boldsymbol{\Delta}\boldsymbol{\Delta}^\top\|\| \\
&\leq 2\|\|\boldsymbol{\Lambda}_0\|\|\epsilon + \epsilon^2
\end{aligned}.$$

Since $\epsilon$ can be chosen arbitrarily small, this establishes strong identifiability of $\boldsymbol{\Sigma}_0$ in $\boldsymbol{\Theta}^*$.

iii) Condition (A2) is exactly assumption N3).

iv) We can always find a $p \times q$ matrix $\boldsymbol{\Lambda}(\hat{\boldsymbol{\theta}})$ of rank $q$, such that $\boldsymbol{M}(\hat{\boldsymbol{\theta}}) = \boldsymbol{\Lambda}(\hat{\boldsymbol{\theta}})\boldsymbol{\Lambda}(\hat{\boldsymbol{\theta}})^\top$. What remains to be shown is that for any such sequence of $\boldsymbol{\Lambda}(\hat{\boldsymbol{\theta}})$, we can find a sequence of orthogonal rotation matrices $\boldsymbol{Q}$ such that $\|\|\boldsymbol{\Lambda}(\hat{\boldsymbol{\theta}})\boldsymbol{Q} - \boldsymbol{\Lambda}_0\|\| \xrightarrow{p} 0$.

From the Orthogonal Procrustes Theorem (see, for example, Golub and Van Loan 2013, Section 6.4.1) we have that for every $\boldsymbol{\Lambda}(\hat{\boldsymbol{\theta}})$, there exists an orthogonal rotation $\boldsymbol{Q}$ such that

$$\|\|\boldsymbol{\Lambda}(\hat{\boldsymbol{\theta}})\boldsymbol{Q} - \boldsymbol{\Lambda}_0\|\|_F^2 = \|\|\boldsymbol{\Lambda}(\hat{\boldsymbol{\theta}})\|\|_F^2 + \|\|\boldsymbol{\Lambda}_0\|\|_F^2 - 2\text{tr}(\boldsymbol{R}), \tag{S10}$$

where $\boldsymbol{R} = \text{diag}(\sigma_1(\boldsymbol{\Lambda}_0^\top\boldsymbol{\Lambda}(\hat{\boldsymbol{\theta}})), \ldots, \sigma_q(\boldsymbol{\Lambda}_0^\top\boldsymbol{\Lambda}(\hat{\boldsymbol{\theta}})))$ is the diagonal matrix of singular values of $\boldsymbol{\Lambda}_0^\top\boldsymbol{\Lambda}(\hat{\boldsymbol{\theta}})$.

Now by (S10), $\|\|\boldsymbol{\Lambda}(\hat{\boldsymbol{\theta}})\|\|_F^2$ converges to $\|\|\boldsymbol{\Lambda}_0\|\|_F^2$ in probability. To show that the singular values $\sigma_i(\boldsymbol{\Lambda}_0^\top\boldsymbol{\Lambda}(\hat{\boldsymbol{\theta}}))$ converge to $\sigma_i(\boldsymbol{\Lambda}_0^\top\boldsymbol{\Lambda}_0)$, note that these singular values are the square roots of the eigenvalues of $\boldsymbol{T}_n = \boldsymbol{\Lambda}_0^\top\boldsymbol{\Lambda}(\hat{\boldsymbol{\theta}})\boldsymbol{\Lambda}(\hat{\boldsymbol{\theta}})^\top\boldsymbol{\Lambda}_0$. Let $\boldsymbol{T} = \boldsymbol{\Lambda}_0^\top\boldsymbol{\Lambda}_0\boldsymbol{\Lambda}_0^\top\boldsymbol{\Lambda}_0$. Then

$$\begin{aligned}
\|\|\boldsymbol{T}_n - \boldsymbol{T}\|\|_2 &= \|\|\boldsymbol{\Lambda}_0(\boldsymbol{\Lambda}(\hat{\boldsymbol{\theta}})\boldsymbol{\Lambda}(\hat{\boldsymbol{\theta}})^\top - \boldsymbol{\Lambda}_0\boldsymbol{\Lambda}_0^\top)\boldsymbol{\Lambda}_0^\top\|\|_2 \\
&\leq \|\|\boldsymbol{\Lambda}_0\|\|_2 \|\|\boldsymbol{\Lambda}(\hat{\boldsymbol{\theta}})\boldsymbol{\Lambda}(\hat{\boldsymbol{\theta}})^\top - \boldsymbol{\Lambda}_0\boldsymbol{\Lambda}_0^\top\|\|_2 \\
&\xrightarrow{p} 0
\end{aligned}$$

By Weyl's inequality (see, for example, Golub and Van Loan 2013, Corollary 8.16),

$$\left|\lambda_i(\boldsymbol{T}_n) - \lambda_i(\boldsymbol{T})\right| \leq \|\|\boldsymbol{T}_n - \boldsymbol{T}\|\|_2 \xrightarrow{p} 0, \quad (i = 1, \ldots, q).$$

Thus the singular values $\sigma_i(\boldsymbol{\Lambda}_0^\top\boldsymbol{\Lambda}(\hat{\boldsymbol{\theta}}))$, which are the square roots of $\lambda_i(\boldsymbol{T}_n)$, converge to $\sigma_i(\boldsymbol{\Lambda}_0^\top\boldsymbol{\Lambda}_0)$, the square roots of $\lambda_i(\boldsymbol{T})$. Therefore

$$\text{tr}(\boldsymbol{R}) \xrightarrow{p} \sum_{i=1}^{q} \sigma_i(\boldsymbol{\Lambda}_0^\top\boldsymbol{\Lambda}_0) = \|\|\boldsymbol{\Lambda}_0\|\|_F^2,$$

15

and thus by (S10) indeed
$$\left\|\left\|\mathbf{\Lambda}(\hat{\boldsymbol{\theta}})\boldsymbol{Q} - \mathbf{\Lambda}_0\right\|\right\|_F^2 \xrightarrow{p} 0\,.$$

□



\end{document}